\documentclass[sn-mathphys]{sn-jnl}

\usepackage{epsfig}
\usepackage{bm}
\usepackage{amsmath,amssymb,amsfonts}
\usepackage{xcolor}
\usepackage{manyfoot}
\usepackage{graphicx}
\usepackage{multirow}
\usepackage{amsthm}
\usepackage{textcomp}

\begin{document}

\title{Helical randomization of magnetized Galactic and galaxy clusters plasmas: from magnetorotational disc dynamo to the Faraday rotation and  synchrotron emission skies}

\author*{\fnm{Alexander} \sur{Bershadskii}}
\email{bershads@gmail.com}

\affil{\orgname{ICAR}, \orgaddress{\city{P.O. Box 31155, Jerusalem}, \postcode{91000}, \country{Israel}}}

\abstract{
  Using results of numerical simulations and Galactic and galaxy clusters observations, it is shown that the transition from deterministic chaos to hard turbulence in the Galactic and galaxy clusters magnetized plasmas occurs via a randomization process. The notion of distributed chaos has been used to describe the randomization process. The randomization can be quantified with the main parameter of the distributed chaos, which in turn can be related to magnetic helicity or its dissipation rate. Spontaneous breaking of local reflectional symmetry (an intrinsic property of chaotic/turbulent motions) generates local helicity even when the global helicity is negligible. It is shown that the magnetic fields can impose their level of randomization on the electron density, Faraday rotation maps, and synchrotron emission. Results of the numerical simulations of the Galactic and galaxy clusters dynamos: the inner disk's ones (based on the magnetorotational instability) and global ones, are in good agreement with this approach, as well as with the results obtained using observations of the Faraday rotation and synchrotron emission skies. }

\maketitle

\section{Introduction}
     
   The observational estimates of the magnetic fields of the Galaxy and galaxy clusters turned out to be much larger than those predicted for the primordial magnetic fields. Therefore different mechanisms of amplification of the seed magnetic fields by the intensive motion of the electrically conducting Galactic and galaxy clusters plasmas were suggested in the literature.  The intensive chaotic/turbulent motion of the Galactic plasmas can be produced, for instance, by the differential rotation of the Galaxy and by the supernova explosions. The special role of a supermassive black hole at the Galactic center in supporting the intensive chaotic/turbulent motion in the inner accretion disk's plasma should also be emphasized.  In the galaxy clusters, chaotic/turbulent motion could be generated, for instance, by cluster mergers, structure formation shocks, active galactic nuclei outflows, and the moving galaxy wakes.  \\
   
   Numerous models (numerical simulations) of the Galactic and galaxy clusters magnetohydrodynamic dynamo were suggested (see, for instance, \citealt{rss,sok,vaz,sub,kor} and references therein). Some recent models will be discussed below in detail. \\
   
   For the weakly compressible central disk, the physical processes at kinetic scales become important and demand fully kinetic dynamo simulations (some of such recent numerical simulations will also be discussed below). \\
   
    The numerical models (simulations) not only allow a better understanding of the relevant physical processes but they become indispensable because of a principal difficulty of measurements (observations) of Galactic and galaxy clusters magnetic fields. Till now technically all observables related to the Galactic and galaxy clusters' magnetic fields are entangled and mixed with other physical characteristics of the magnetized plasmas (electron density, for instance).  \\
    
    Another principal difficulty in this area is the absence of a relevant theory of the chaotic/turbulent processes. The scaling approach, widely used for data interpretation,  demands a rather wide range of scales to validate its applicability which is scarcely achievable in practice.   \\
   
    In the absence of such a theory, the notion of smoothness can be used to classify (quantify) chaotic/turbulent dynamical regimes. For this purpose, one can use spectral analyses because the stretched exponential spectra
\begin{equation}
E(k) \propto \exp-(k/k_{\beta})^{\beta} 
\end{equation}
are typical for smooth chaotic dynamics. Here $k$ is the wavenumber and $1 \geq \beta > 0$. 

The particular value $\beta =1$ typically characterizes the deterministic chaos (see, for instance, \citealt{fm, swinney1, mm1, mm2, mm3, kds} and references therein):
\begin{equation}
E(k) \propto \exp(-k/k_c).  
\end{equation}

 For $1 > \beta$ the smooth chaotic dynamics is not deterministic (this dynamics will be called `distributed chaos' and the term will be clarified below). The term ``soft turbulence'' suggested in a paper (\citealt{wu}) could be also appropriate.\\

 The value of the $\beta$ can be used as a measure of randomization: the further the value of  $\beta$ is from the $\beta =1$ corresponding to deterministic chaos, the more significant the randomization. The small value of $\beta$ can be considered as a precursor of hard turbulence (\citealt{wu}). For non-smooth dynamics (the hard turbulence \citealt{wu}) the power-law (scaling) spectra are typical.\\ 

\begin{figure} \vspace{-2.2cm}\centering \hspace{-1.1cm}
\epsfig{width=.65\textwidth,file=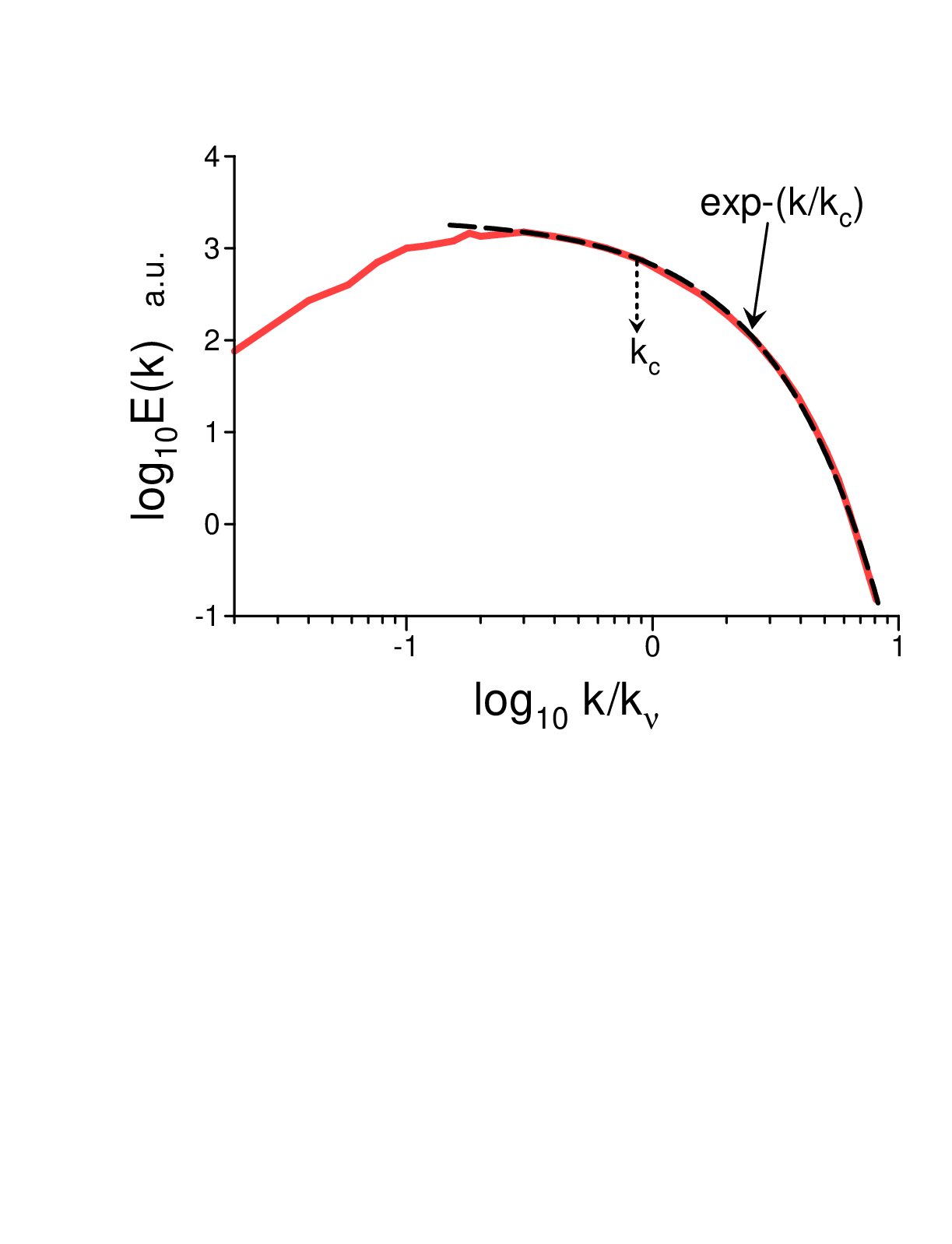} \vspace{-4.6cm}
\caption{Magnetic energy spectrum at the saturated stage of a small-scale dynamo for a sufficiently large magnetic Reynolds number $Re_m =3000$ and small Reynolds and Mach numbers $Re =100$, $M = 0.122$} 
\end{figure}
    
  Figure 1, for instance,  shows the magnetic energy spectrum obtained in a recent direct numerical simulation of a small-scale magnetohydrodynamic dynamo adapted to galaxy plasmas (\citealt{bzs})  (see also Fig. 14 below). The spectral data were taken from Fig. 4 of the paper by \citealp{bzs}.\\
  
  The authors of the paper considered a weakly compressible turbulence with an isothermal equation
of state with the pressure $p=\rho c_s^2$, where  $\rho$ is the density and $c_s$ is the constant speed of sound. The dynamics of the magnetic field ${\bf B} =\nabla\times{\bf A}$ (where ${\bf A}$ is the magnetic vector potential),  the velocity ${\bf u}$, and the density $\rho$ was described by the system of equations:
\begin{equation}
\frac{\partial {\bf A}}{\partial t}={\bf u}\times {\bf B}+\eta\nabla^2 {\bf A},
\end{equation}
\begin{equation}
\frac{D{\bf u}}{D t}=-c_s^2\nabla\ln\rho+
\frac{1}{\rho}\left[{\bf J}\times{\bf B}+\nabla\cdot(2\rho\nu {\bf S})\right] +{\bf f},
\end{equation}
\begin{equation}
\frac{D\ln\rho}{D t}=-\nabla\cdot {\bf u},
\end{equation}
where the current density ${\bf J}=\nabla \times {\bf B}/\mu_0$ ($\mu_0$ is the vacuum permeability), $\nu$ is viscosity, ${\bf S}$ is the rate-of-strain tensor,
and ${\bf f}$ is a random nonhelical forcing function.\\

  Figure 1 shows the magnetic energy spectrum at the saturated stage of the small-scale dynamo for a sufficiently large magnetic Reynolds number $Re_m =3000$ and a small Reynolds and Mach numbers $Re =100$, $M = 0.122$ (magnetic Prandtl number $Pr_m =30$). In this case, one can expect deterministic chaos in the magnetic field generated by the dynamo. And indeed the dashed curve in Fig. 1 indicates the best fit by the exponential spectrum Eq. (2) typical for the deterministic chaos. The dotted arrow indicates position of $k_c/k_{\nu}$ where $k_{\nu} =( \varepsilon/{\nu}^3)^{1/4}$ is the Kolmogorov or viscous dissipation scale.\\

   In recent papers (\citealt{sf,seta}) results of analogous simulations were reported. But in this case, the Mach number $M=10$ was reached. Figure 2 shows the magnetic energy spectra obtained in these simulations. The spectral data were taken from Fig. C4 of the Ref. (\citealt{seta}).  One can see that for $M=0.1$ the dashed curve indicates the best fit by the exponential spectrum Eq. (2) typical for the deterministic chaos (cf Fig. 1), whereas for $M=10$ the dashed curve indicates the best fit by a stretched exponential spectrum Eq. (1) with $\beta = 1/2$, i.e. the distributed chaos. 

\begin{figure} \vspace{-1.8cm}\centering \hspace{-1.1cm}
\epsfig{width=.7\textwidth,file=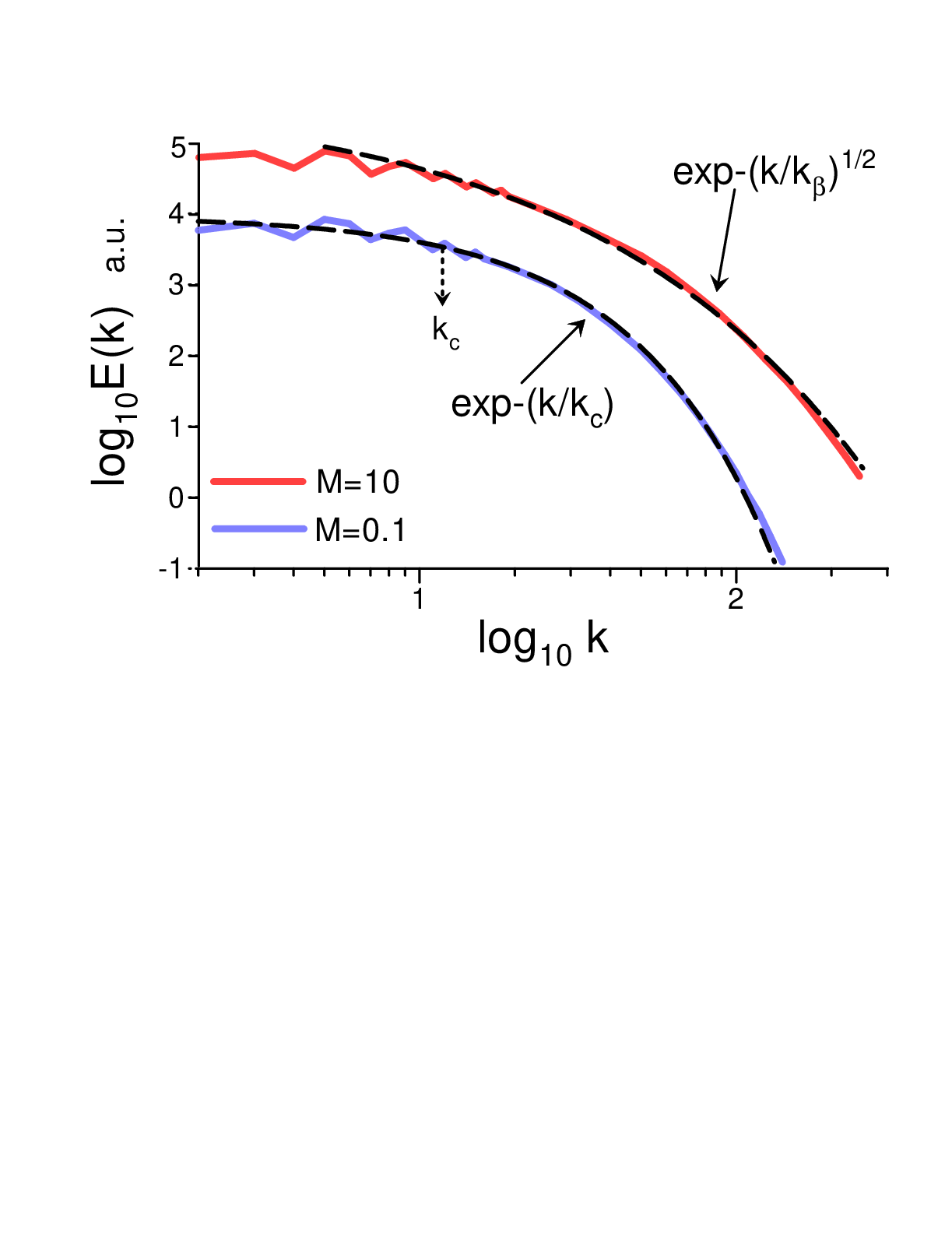} \vspace{-5.45cm}
\caption{Magnetic energy spectrum at the saturated stage of a small-scale dynamo for $M=0.1$ and $M=10$.} 
\end{figure}
    
    It will be shown in the next Section 2 that the value of $\beta = 1/2$ corresponds to the distributed chaos dominated by the magnetic helicity and the results of a fully kinetic 3D numerical simulation of a magnetorotational dynamo in the inner galactic disc will be interpreted in this vein (for the case of zero net magnetic helicity the spontaneous breaking of local reflectional symmetry will be used to justify this approach).  Then in Section 3 a magneto-inertial range of scales dominated by magnetic helicity will be introduced in the frames of the Kolmogorov phenomenology (which is characterized by different values of $\beta$, see also \citealt{ber4}). Section 4 will compare the theoretical considerations to the results of numerical simulations of the global galactic magnetic field and observations of the Galaxy's Faraday rotation sky. In Section 5 the synchrotron emission as a tracer of the Galactic magnetic field will be considered. In Section 6 galaxy clusters' intracluster magnetized plasma and its magnetic field will be considered using analogous tools.
   
\section{Distributed chaos and magnetic helicity}  
 
 \subsection{Magnetic helicity}   
 
 The ideal magnetohydrodynamics has three quadratic (fundamental) invariants: total energy, cross and magnetic helicity (\citealt{mt}).  The robustness of magnetic helicity conservation depends on the magnetic Reynolds number value $Re_m$. Most astrophysical systems (including the galaxies) have very large magnetic Reynolds numbers (see, for instance, a recent paper by \citealt{zv} and references therein).\\
  
  The average magnetic helicity density can be defined as
\begin{equation}
 h_m = \langle {\bf a} {\bf b} \rangle   
\end{equation}
where ${\bf b} = [{\nabla \times \bf a}]$ is the fluctuating magnetic field, ${\bf a}$ is the vector potential, and $\langle ... \rangle$ denotes a spatial average. The fluctuating magnetic field ${\bf b}$ and magnetic potential ${\bf a}$ have zero means. \\

  The invariance of $h_m$ is not valid in a uniform mean magnetic field ${\bf B_0}$. To deal with this problem a generalized average magnetic helicity density $\hat{h}$ was considered in the paper (\citealt{mg})
 \begin{equation}
 \hat{h} = h_m + 2{\bf B_0}\cdot \langle {\bf A}  \rangle 
\end{equation}
where ${\bf B} = {\bf B_0} + {\bf b}$, ${\bf A} = {\bf A_0} +{\bf a}$. It was shown that in the ideal magnetohydrodynamics (\citealt{mg}) (see also \citealt{shebalin})
\begin{equation}
 \frac{d \hat{h}}{d t} =  0
\end{equation}
That means invariance of the generalized average magnetic helicity  density $\hat{h} $ in a uniform mean magnetic field.\\

\subsection{Distributed chaos driven by magnetic helicity} 
   
    One can see in the Introduction (Fig. 2) that a change of physical parameters (in this case the Mach number) can result in a transition from deterministic chaos to distributed one (randomization).
     
     Let us understand the transition. The change of physical parameters results in fluctuations of the characteristic scale $k_c$ in equation (2). One can take this phenomenon into account with the help of an ensemble averaging 
\begin{equation}
E(k) \propto \int_0^{\infty} P(k_c) \exp -(k/k_c)dk_c 
\end{equation}  
where the random fluctuations of $k_c$ are described by a probability {\it distribution} $P(k_c)$. This is the reason why the non-deterministic smooth chaotic dynamics has been called `distributed chaos'.\\

 One can use the dimensional considerations and the scaling relationship between $k_c$ and the characteristic value of the magnetic field $B_c$
\begin{equation}
B_c \propto |h_m|^{1/2} k_c^{1/2}   
\end{equation}
to find $P(k_c)$ for the magnetic field dynamics dominated by the magnetic helicity. \\

  For this purpose let $B_c$ be half-normally distributed $P(B_c) \propto \exp- (B_c^2/2\sigma^2)$ (\citealt{my}). The half-normal distribution is a normal distribution with zero mean truncated to only have nonzero probability density for positive values of its argument. For a normal random variable $B$, for instance, the variable $B_c = |B|$ has a half-normal distribution (\citealt{jkb}). \\
  
    From Eq. (10) we then obtain 
\begin{equation}
P(k_c) \propto k_c^{-1/2} \exp-(k_c/4k_{\beta})  
\end{equation}
i.e. the chi-squared distribution (here $k_{\beta}$ is a new constant). \\

   Substituting Eq. (11) into Eq. (9) we obtain
\begin{equation}
E(k) \propto \exp-(k/k_{\beta})^{1/2}  
\end{equation}

  The average magnetic helicity density $h_m$ was used in the estimate Eq. (10) as an invariant. Taking into account that there are no physical processes effectively transferring the magnetic helicity to the resistive scales one can consider it as an adiabatic invariant (in comparison with the effectively dissipating magnetic and total energy) also in a weakly dissipative case (see for instance, \citealt{zv}).

\subsection{Spontaneous breaking of local reflectional symmetry}

  In a special case of global (net) reflectional symmetry, the global (net) magnetic helicity is equal to zero. In contrast, the point-wise magnetic helicity is not (because of the spontaneous breaking of the local reflectional symmetry). This is an inherent property of chaotic/turbulent flows. The emergence of the blobs with non-zero magnetic/kinetic helicity accompanies the spontaneous local symmetry breaking (\citealt{mt,moff1,moff2,kerr,hk,lt}). The magnetic surfaces are defined by the boundary conditions: ${\bf b_n}\cdot {\bf n}=0$, where ${\bf n}$ is a unit normal to the boundary of a magnetic blob. The sign-defined magnetic helicity of any magnetic blob is an adiabatic invariant (\citealt{mt}) (see also above). The magnetic blobs can be numbered with $H_j^{\pm}$ denoting the helicity of the j-blob and (`+'  or `-') denoting the blob's helicity sign:
\begin{equation}
 H_j^{\pm} = \int_{V_j} ({\bf a} ({\bf x},t) \cdot  {\bf b} ({\bf x},t)) ~ d{\bf x} 
\end{equation}  
  
  Then we can consider the adiabatic invariant 
\begin{equation}
{\rm I^{\pm}} = \lim_{V \rightarrow  \infty} \frac{1}{V} \sum_j H_{j}^{\pm}   
\end{equation}
 The summation here is over the blobs with a certain sign only (`+' or `-'), and $V$ is the entire volume of the blobs participating in the summation.  \\

  The adiabatic invariant ${\rm I^{\pm}}$ can be used instead of $h_m$ in the estimate Eq. (10) for the case of the local reflectional symmetry breaking
\begin{equation}
B_c \propto |{\rm I^{\pm}}|^{1/2} k_c^{1/2}   
\end{equation}
and the stretched exponential spectrum Eq. (12) can be obtained for this case as well.\\

\begin{figure} \vspace{-2.5cm}\centering 
\epsfig{width=.7\textwidth,file=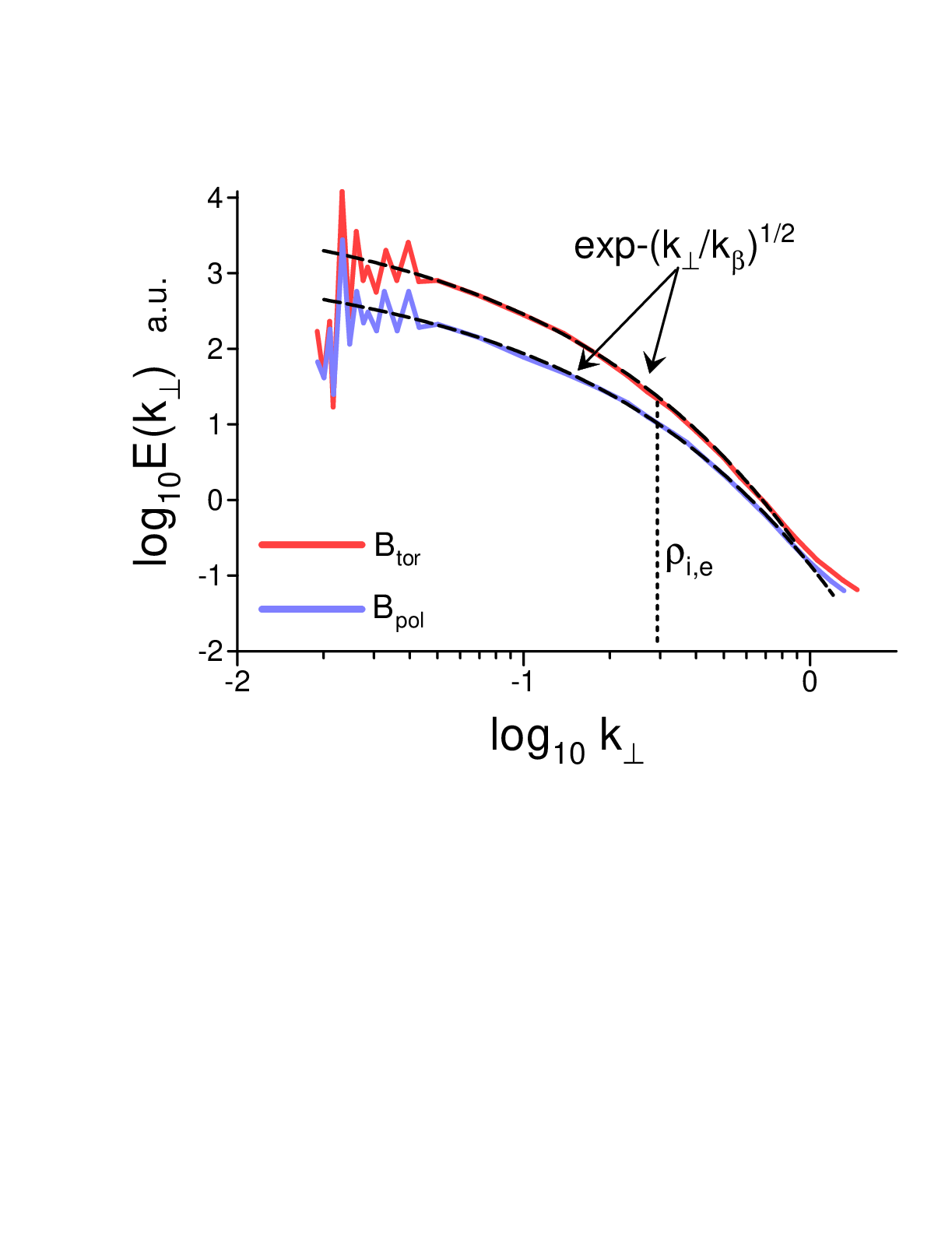} \vspace{-4.55cm}
\caption{Magnetic energy spectra (vs. the perpendicular to the magnetic field wavenumber $k_{\perp}$) obtained in a recent 3D fully kinetic particle-in-cell simulation of a magnetorotational disc dynamo in collisionless pair plasma at a quasi-stationary non-linear stage.} 
\end{figure}

  If a non-zero mean magnetic field ${\bf B}_0$ is not negligible the average magnetic helicity density $h_m$ could be replaced in the estimate Eq. (10) by the generalized averaged magnetic helicity density $\hat{h}$ Eq. (7) . Since the generalized averaged magnetic helicity density $\hat{h}$ has the same dimensionality as $h_m$ the magnetic energy spectrum will be the same stretched exponential Eq. (12) (analogously can be considered the case of the local reflectional symmetry breaking).\\
 
  The stretched exponential spectrum Eq. (12) already appeared in Fig. 2 at the saturated stage of a small-scale dynamo for the Mach number $M=10$ (see also Section 5). \\
  
    Figure 3 shows the magnetic energy spectra (vs. the perpendicular to the magnetic field wavenumber $k_{\perp}$) obtained in a recent fully kinetic particle-in-cell simulation of a magnetorotational disc dynamo in collisionless pair plasma at a quasi-stationary non-linear stage. The spectral data were taken from Fig 2a of a recent paper (\citealt{bac}).\\
    
     This numerical simulation was related to the magnetorotational disk dynamo (based on the magnetorotational instability) in the high-energy surrounding of a supermassive black hole at the Galactic center (see Introduction). The kinetic scales can play a significant role in the physics of the collisionless plasmas therefore the fully kinetic numerical simulations are very important (about conservation of magnetic helicity at kinetic scales see, for instance \citealt{sch1,pm,ber4}). \\
    
    The dashed curves indicate the best fit by the stretched exponential Eq. (12) both for the toroidal and poloidal components of the magnetic field. The vertical dotted line indicates the positions of the ion and electron gyroradii which coincide in this case. 
  
\section{Magneto-inertial range of scales}   

   An inertial range of scales is expected in hydrodynamic turbulence for the high Reynolds numbers. In the inertial range the energy spectra are supposed to be depending on the energy dissipation rate $\varepsilon$ only (\citealt{my}). \\
    
    A magneto-inertial range of scales has been introduced recently in plasma dynamics for the magnetohydrodynamic and kinetic scales (\citealt{ber4}).  In this range of scales, two parameters: the total energy dissipation rate $\varepsilon$ (see above) and the magnetic helicity dissipation rate $\varepsilon_h$ (or dissipation rate of its modification $I^{\pm}$ in the case of global reflectional symmetry) govern the magnetic field dynamics. \\
    
    There is an analogy of this approach to the Corrsin-Obukhov approach to the passive scalar in the turbulent flows where two governing parameters: the energy dissipation rate and the passive scalar dissipation rate, determined the so-called inertial-convective range (\citealt{my}) (see also \citealt{bs1}). \\
    
    According to this analogy, we can replace the estimate Eq. (10) with the estimate
\begin{equation}
 B_c \propto \varepsilon_h^{1/2} ~\varepsilon^{-1/6}~k_c^{1/6}  
\end{equation}
 for the magneto-inertial range.\\
 
     The estimates Eq. (10), (15), and (16) can be generalized
\begin{equation}
 B_c \propto k_c^{\alpha}   
\end{equation}
  
  At large $k_c$ the stretched exponential representation of the distributed chaos spectra 
\begin{equation}
 \int_0^{\infty}  P(k_c) \exp -(k/k_c)dk_c \propto \exp-(k/k_{\beta})^{\beta} 
\end{equation}  
can be employed for finding the probability distribution $P(k_c)$ (\citealt{jon})
\begin{equation}
P(k_c) \propto k_c^{-1 + \beta/[2(1-\beta)]}~\exp(-\gamma k_c^{\beta/(1-\beta)}) 
\end{equation}   
  
  Using this asymptote a relationship between the parameters $\alpha$ and $\beta$ can be derived (using some simple algebra) from the Eqs. (17) and (19) in the case of the half-normally distributed $B_c$ 
\begin{equation}
\beta = \frac{2\alpha}{1+2\alpha}  
\end{equation}

 Then for $\alpha =1/6$ (see Eq. (16)), one can find from Eq. (20) that  
\begin{equation}
 E(k) \propto \exp-(k/k_{\beta})^{1/4} , 
\end{equation}
 
\begin{figure} \vspace{-2.1cm}\centering 
\epsfig{width=.65\textwidth,file=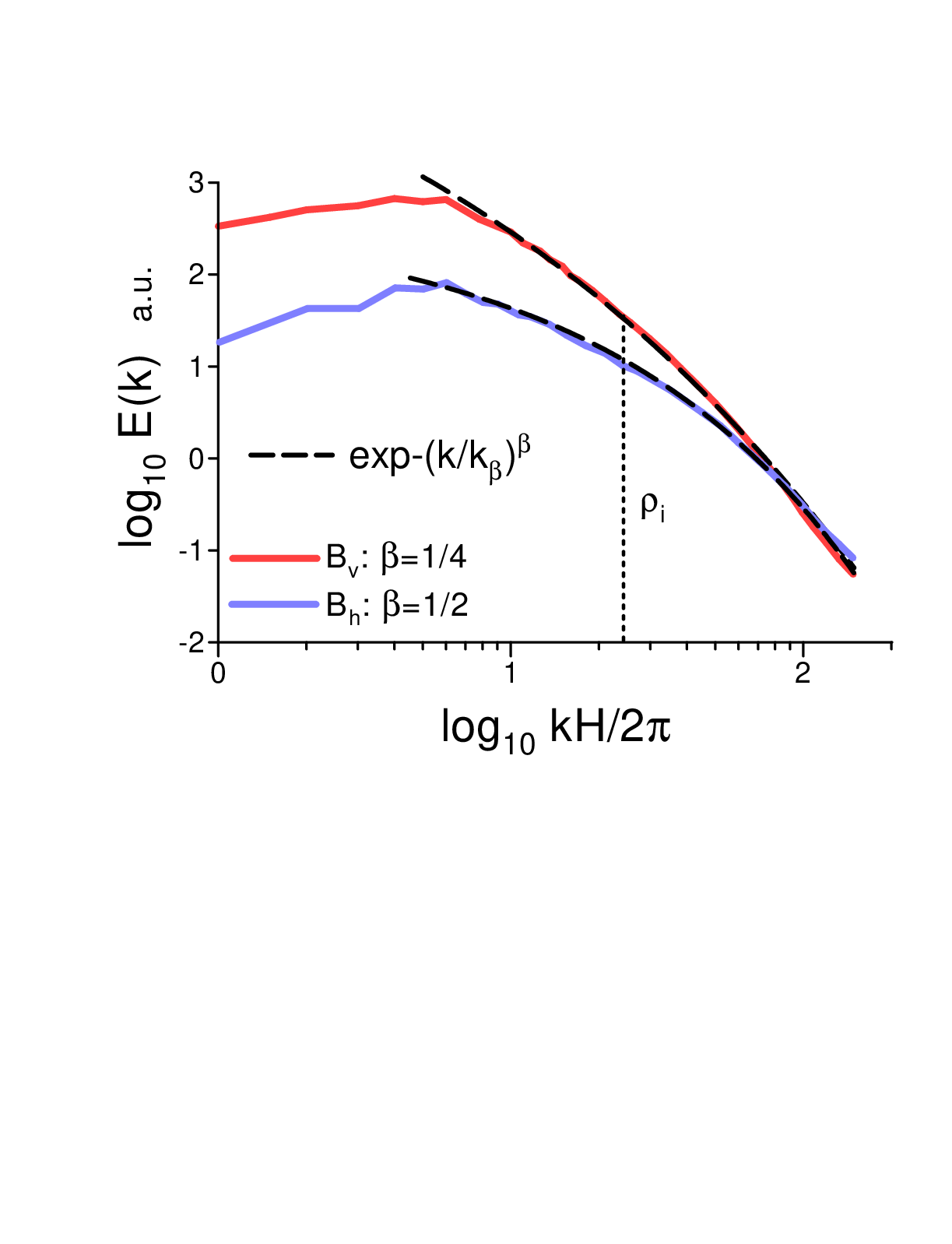} \vspace{-4.2cm}
\caption{Magnetic energy spectra obtained in a 3D kinetic numerical simulation of magnetorotational dynamo based on the magnetorotational instability in a collisionless accretion disk at the saturated stage.} 
\end{figure}
  
for the magneto-inertial range of scales.\\  

    Figure 4 shows the magnetic energy spectra obtained in a 3D kinetic numerical simulation of magnetorotational dynamo based on the magnetorotational instability in a collisionless accretion disk (cf Fig. 3). The spectral data were taken from Fig. 5a of a paper (\citealt{kunz}). The wavenumber $k$ was normalized by the disk height $H$. The magnetic Reynolds number was rather large $Re_m =37500$, ${\bf B}_ v$ and ${\bf B}_h$ are vertical and horizontal magnetic field components, respectively.\\
    
    The dashed curves indicate the spectral law Eq. (12) for ${\bf B}_h$ component and Eq. (21) for ${\bf B}_v$ component (an anisotropy of the level of randomization). The vertical dotted line indicates the position of the ion gyroradius $\rho_i$. \\
    
    Figure 5 shows the power spectrum of the plasma density fluctuations at the saturated stage. The spectral data were taken from Fig 5b of the paper (\citealt{kunz}). The dashed curve indicates the spectral law Eq. (12). One can see that the horizontal component of the magnetic field imposes its level of the randomization $\beta =1/2$ on the plasma density fluctuations in this case (see also next Section). 
    
\section{Global galactic magnetic field and Faraday rotation sky}   

\subsection{Numerical simulations}
  
  Let us start with numerical simulations. In the Introduction, we have already discussed two simplified numerical simulations of the Galactic magnetic field (see Figs. 1 and 2). In a recent paper (\citealt{nto}) generation (amplification) of the global magnetic field of a Milky-Way-like galaxy by the $\alpha-\omega$ dynamo was numerically simulated using a system of equations:
\begin{eqnarray}
\frac{\partial\rho}{\partial t} +\nabla(\rho{\bf u})& =& 0 \\
\frac{\partial\rho{\bf u}}{\partial t} + \nabla\cdot\left(\rho\bf{u}\bf{u}-\bf{B}\bf{B}\right) + \nabla P_{tot} &=& -\rho\nabla\phi  \\
\frac{\partial E_{tot}}{\partial t} +\nabla\left[~(E_{tot}+P_{tot})\bf{u} -(\bf{u}\cdot\bf{B})\cdot\bf{B}\right] &=& -{\bf u}\cdot\nabla\phi -\rho\Lambda + \Gamma \\
\frac{\partial\bf{B}}{\partial t} -\nabla\times(\bf{u}\times\bf{B}) &=& 0  \\
\nabla\cdot\bf{B} &=& 0 
\end{eqnarray}
where  $\phi$ is the gravitational potential, $\Gamma=\Gamma(\rho,T)$ and  $\Lambda=\Lambda(\rho,T)$ (as functions of density and temperature) are the heating and cooling rates of the plasma, the total pressure:
\begin{equation}
    P_{tot} = p + \frac{\bf{B}\cdot\bf{B}}{2}
\end{equation}
and the total energy 
\begin{equation}
    E_{tot} = E_{int} + \rho\frac{\bf{u}\cdot\bf{u}}{2} +  + \frac{\bf{B}\cdot\bf{B}}{2}
\end{equation}
where the internal energy of the fluid is $E_{int}$. The equation of state for the plasma was taken as $P=(\gamma-1)E_{int}$. \\ 

\begin{figure} \vspace{-2cm}\centering 
\epsfig{width=.65\textwidth,file=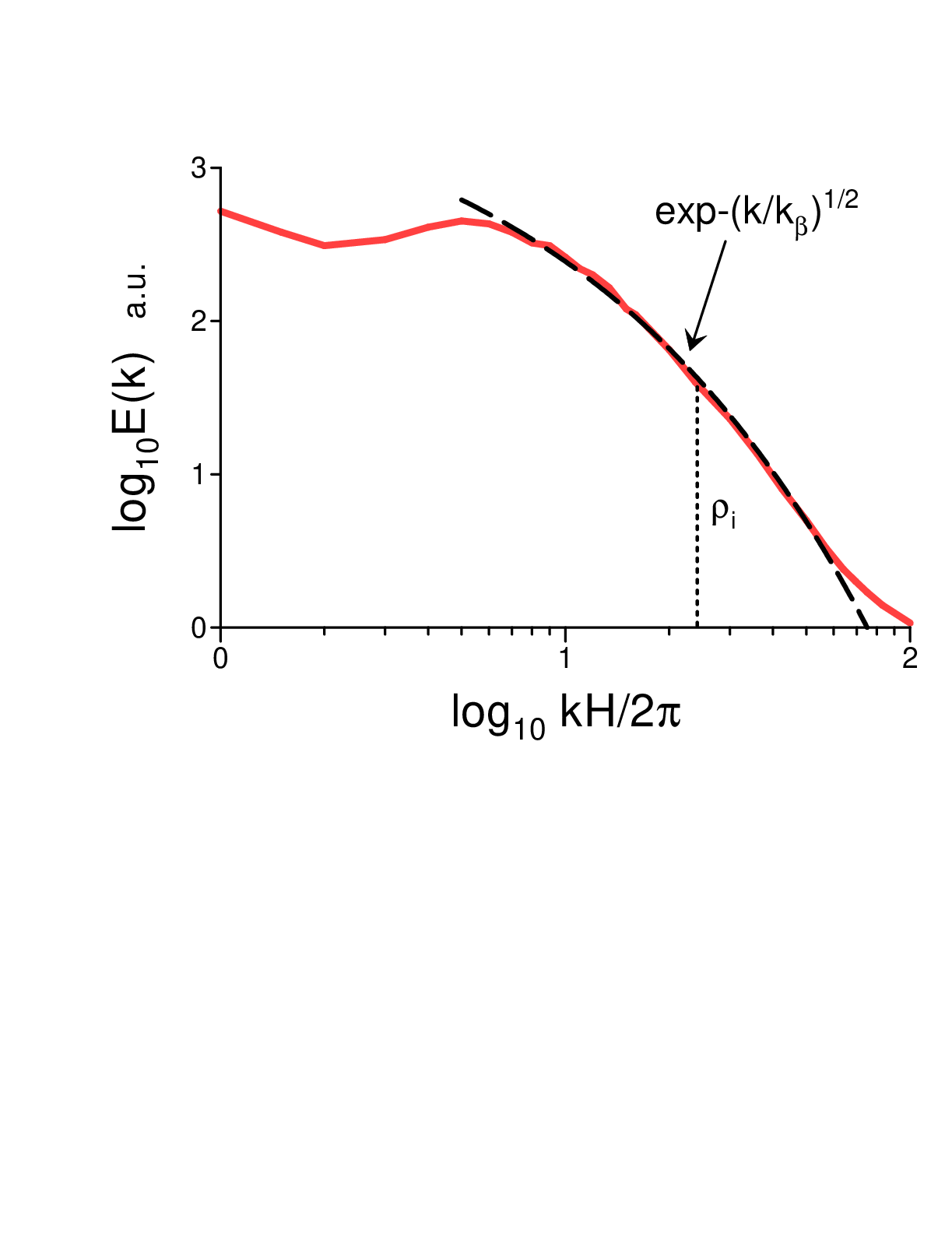} \vspace{-4.3cm}
\caption{As in Fig. 4 but for density fluctuations.} 
\end{figure}
\begin{figure} \vspace{-2.1cm}\centering 
\epsfig{width=.7\textwidth,file=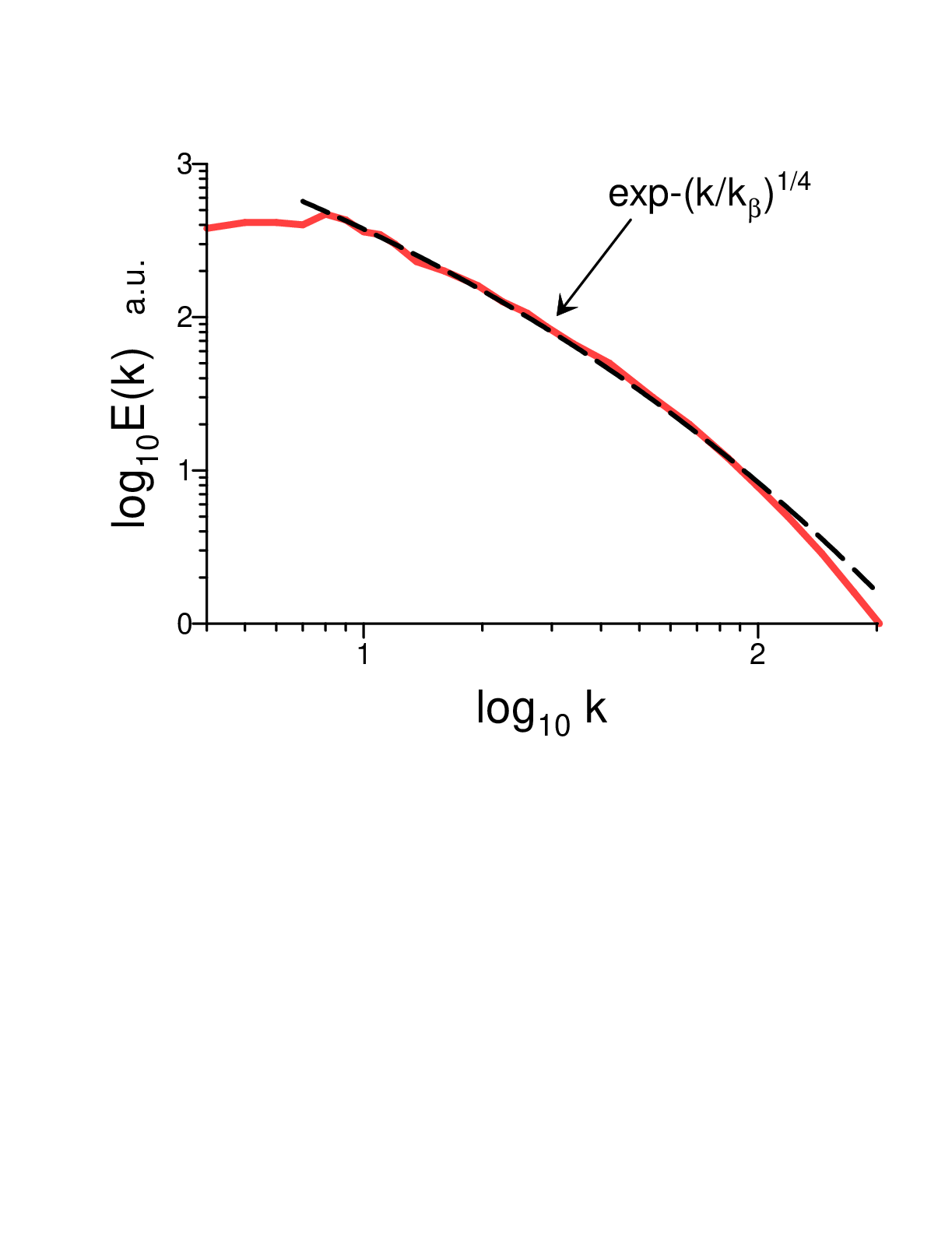} \vspace{-4.8cm}
\caption{Power spectrum of the chaotic/turbulent galactic magnetic field fluctuations (numerical simulations). } 
\end{figure}

  The model also included stars, stellar feedback, and a dark matter halo.  The configuration of hydrodynamical fluid, dark matter, and stellar particles was chosen to simulate a Milky-Way-like galaxy. A weak toroidal magnetic field was taken as a seed initial condition.\\ 
  
  In the simulation, turbulence is generated by the differential rotation of the galaxy and by the supernova explosions (supernova explosions produce a considerable amount of energy which then is transferred to the surrounding interstellar medium).\\

  At a certain stage of the dynamo evolution, the magnetic field was smoothed at a given scale to separate the mean field and chaotic/turbulent residual fluctuations.  \\
  
  Figure 6 shows the magnetic energy spectrum of the chaotic/turbulent magnetic field fluctuations. The dashed curve indicates the best fit by the stretched exponential Eq. (21).\\

\subsection{Observations}

\begin{figure} \vspace{-2.2cm}\centering 
\epsfig{width=.7\textwidth,file=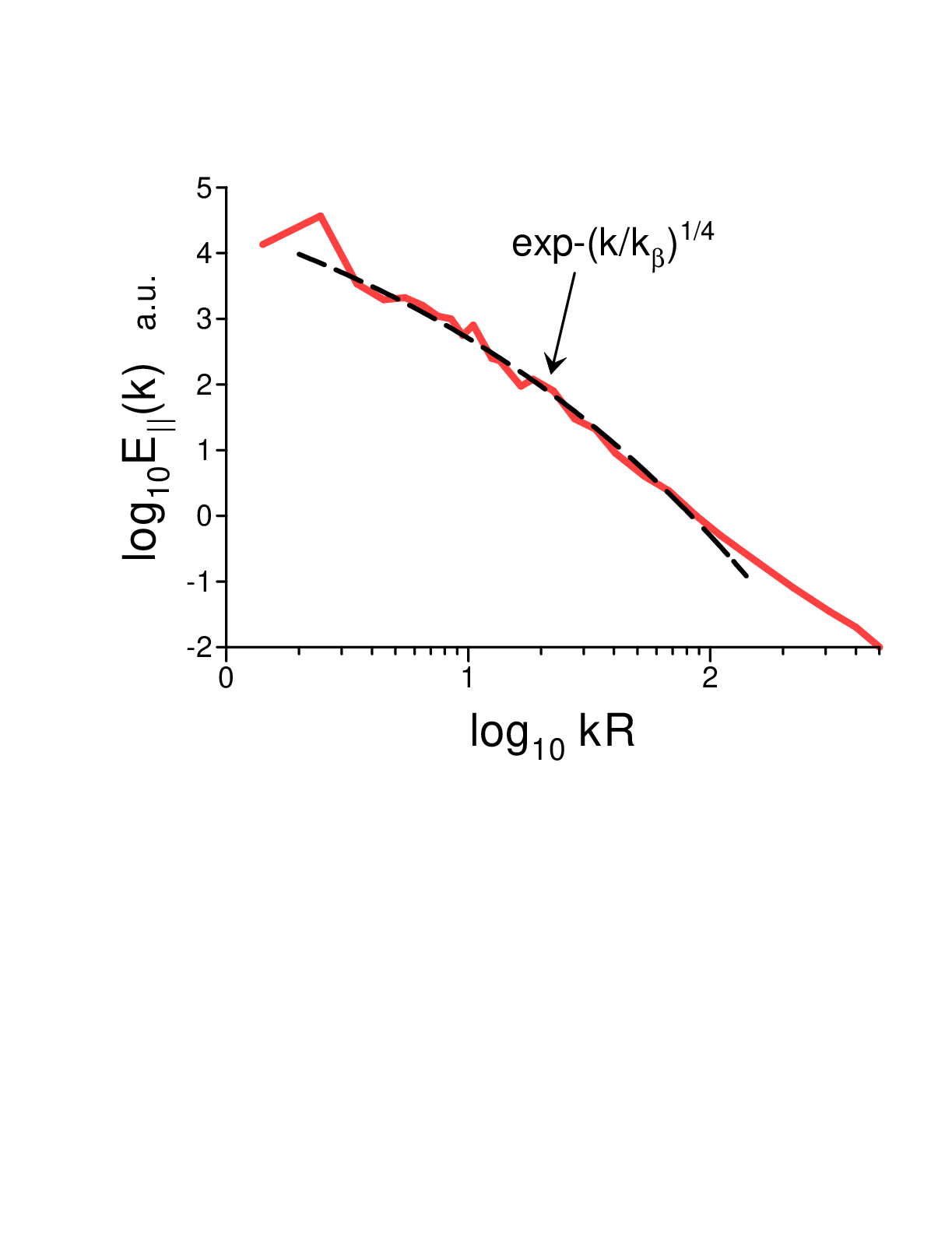} \vspace{-4.6cm}
\caption{Power spectrum of the average $B_{\mathrm{\parallel}}$ sky map (inferred from the Farady rotation sky).} 
\end{figure}
\begin{figure} \vspace{-0.4cm}\centering 
\epsfig{width=.7\textwidth,file=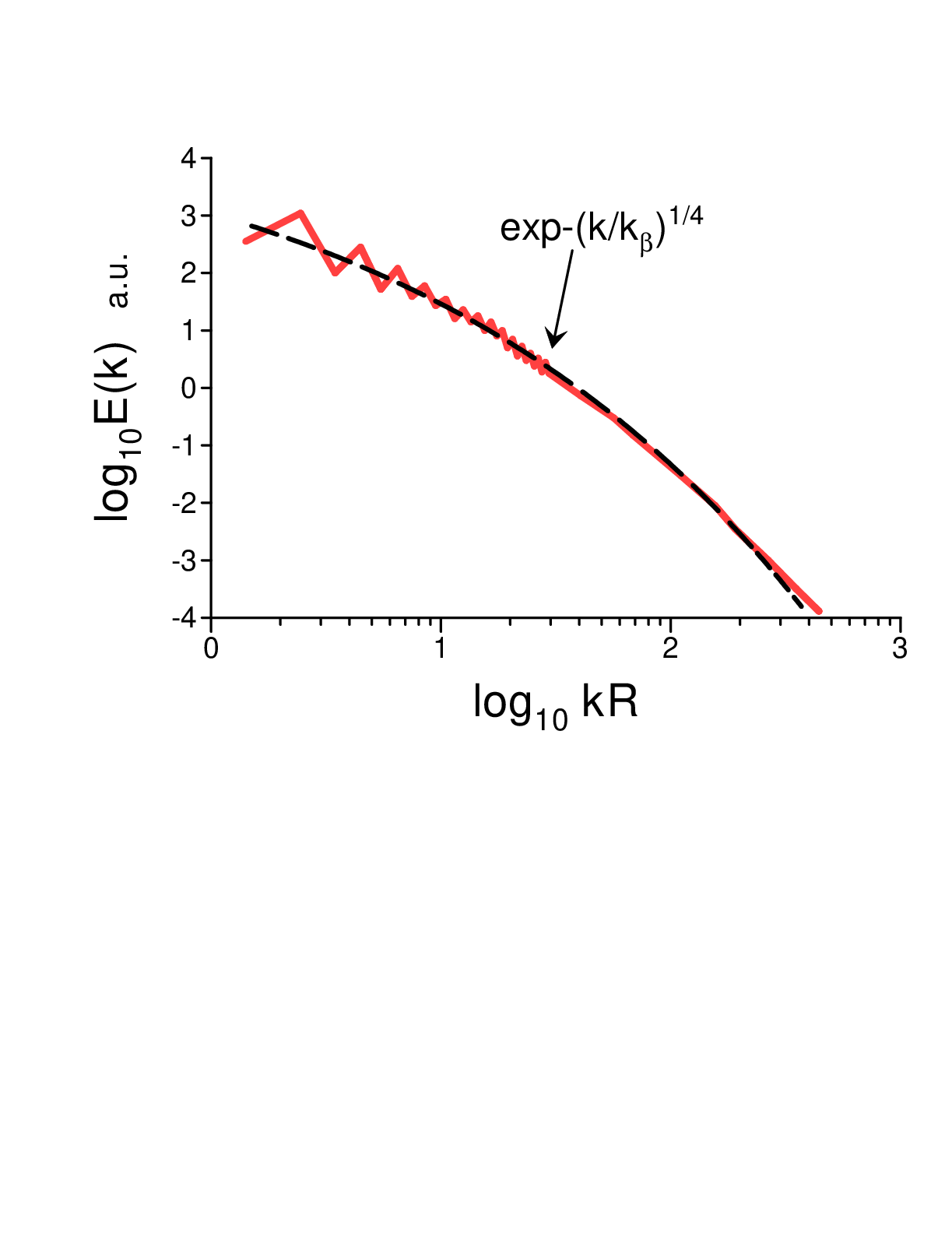} \vspace{-4.8cm}
\caption{Power spectrum of electron dispersion measure sky map (inferred from the Farady rotation sky).} 
\end{figure}

  Now let us turn to observational results related to the Galactic magnetic field. The Faraday effect represents a rotation of the polarization
position angle which propagates through magnetized plasma and entangles information on the plasma electron density $n_e$ with the line-of-sight (LoS) component of the magnetic field $B_{\mathrm{\parallel}}$  into the observable Faraday depth:
\begin{equation}
\phi  = \frac{e^3}{2\pi m_e^2 c^4}\int_{\mathrm{LoS}} dl\, n_{\mathrm{e}} B_{\mathrm{\parallel}} 
\end{equation}
The integral Eq. (29) also determines the rotation measure (RM) often used for the data interpretation (see a recent paper \citealt{hut} and next Section). \\
 
   In the paper (\citealt{hut}) results of disentangling an all-sky map of the average Galactic $B_{\mathrm{\parallel}}$ from the Faraday effect were reported. Several additional tracers of the $n_e$ (a $H-\alpha$ map \citealt{fin}, pulsar data \citealt{man}, the free-free map of the Planck survey \citealt{Planck}, and extra-Galactic Faraday data \citealt{eak}) were used for this purpose and an all-sky map of the electron dispersion measure (the integrated electron density) was also constructed. \\
     
     Figures 7 and 8 show the power spectra of the average $B_{\mathrm{\parallel}}$ and electron dispersion measure respectively. The spectral data were taken from Figs. 9a and 9b of the paper (\citealt{hut}). In the original figures presented in the paper (\citealt{hut}) the spectra are shown vs the spherical harmonic degree $l$. The spherical harmonic degree $l$ can be related to the wave number $k = \sqrt{l(l+1)}/R$, where $R$ is an effective radius of the Galactic map.\\
     
\begin{figure} \vspace{-2cm}\centering 
\epsfig{width=.65\textwidth,file=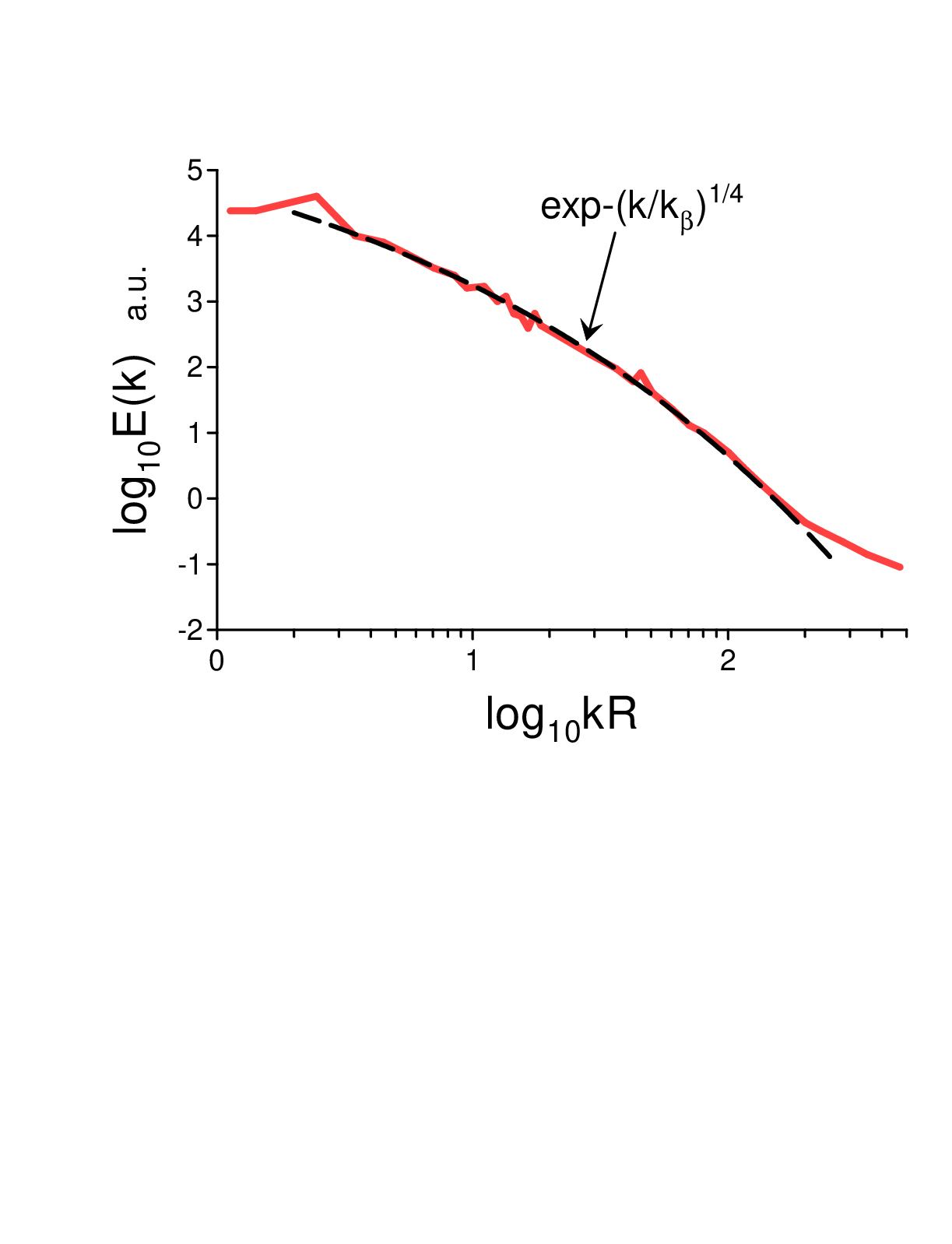} \vspace{-4.3cm}
\caption{Power spectrum of the Galactic Faraday rotation map.} 
\end{figure}

     The dashed curves in Figs. 7 and 8 indicate the best fit by the stretched exponential spectral law Eq. (21). One can see that in this case, the magnetic field imposes its degree of randomization on the electron dispersion measure.\\
     
     Now it is not surprising that the magnetic field also imposes its degree of randomization on the Faraday map itself (see also Section 6). Figure 9 shows the spectrum of the Faraday map. The spectral data were taken from Fig. 6  (reconstruction II) of a paper (\citealt{he}). The dashed curve in Fig. 9 indicates the best fit by the stretched exponential spectral law Eq. (21).

\section{Synchrotron emission as a tracer of magnetic field}

  The astroparticles (electrons and positrons) can produce synchrotron emission due to their gyration in the magnetic field. The diffuse polarized synchrotron emission can be an effective tracer of the magnetic field in the magnetized chaotic/turbulent non-thermal plasma (see, for instance, Ref. \citealt{rob} and references therein). It should also be noted that this emission is one of the main polarized foregrounds for the cosmic microwave background radiation (CMB) and its successful removal from the observed CMB is important for clean cosmological measurements. Therefore, understanding its spectral properties is rather important.\\

  In a Ref. (\citealt{zs}) a spin-2 decomposition of the polarization tensor (vector) was suggested. This technique is applied mainly for analysis of the cosmic microwave background radiation. In the Ref. (\citealt{rob}) it was applied to analyze the diffuse polarized synchrotron emission in the magnetized interstellar plasma. In this technique, the polarization tensor (vector) is decomposed using two rotationally invariant scalar (E) and pseudo-scalar (B) quantities. The B-mode has magnetic-type parity. \\ 

  \citealt{rob} used a polarimetric survey of the sky obtained with the Parkes Radio Telescope (S-band receiver at 2.3 GHz) to analyze the synchrotron emission of certain Galactic regions. In the analyzed Galactic regions, the magnetic field perturbations were trigged by the clusters of stars related to the Orion–Eridanus superbubble, and the shell-like structure in a region denoted with its Galactic coordinates G353-34.\\
  
\begin{figure} \vspace{-2cm}\centering 
\epsfig{width=.65\textwidth,file=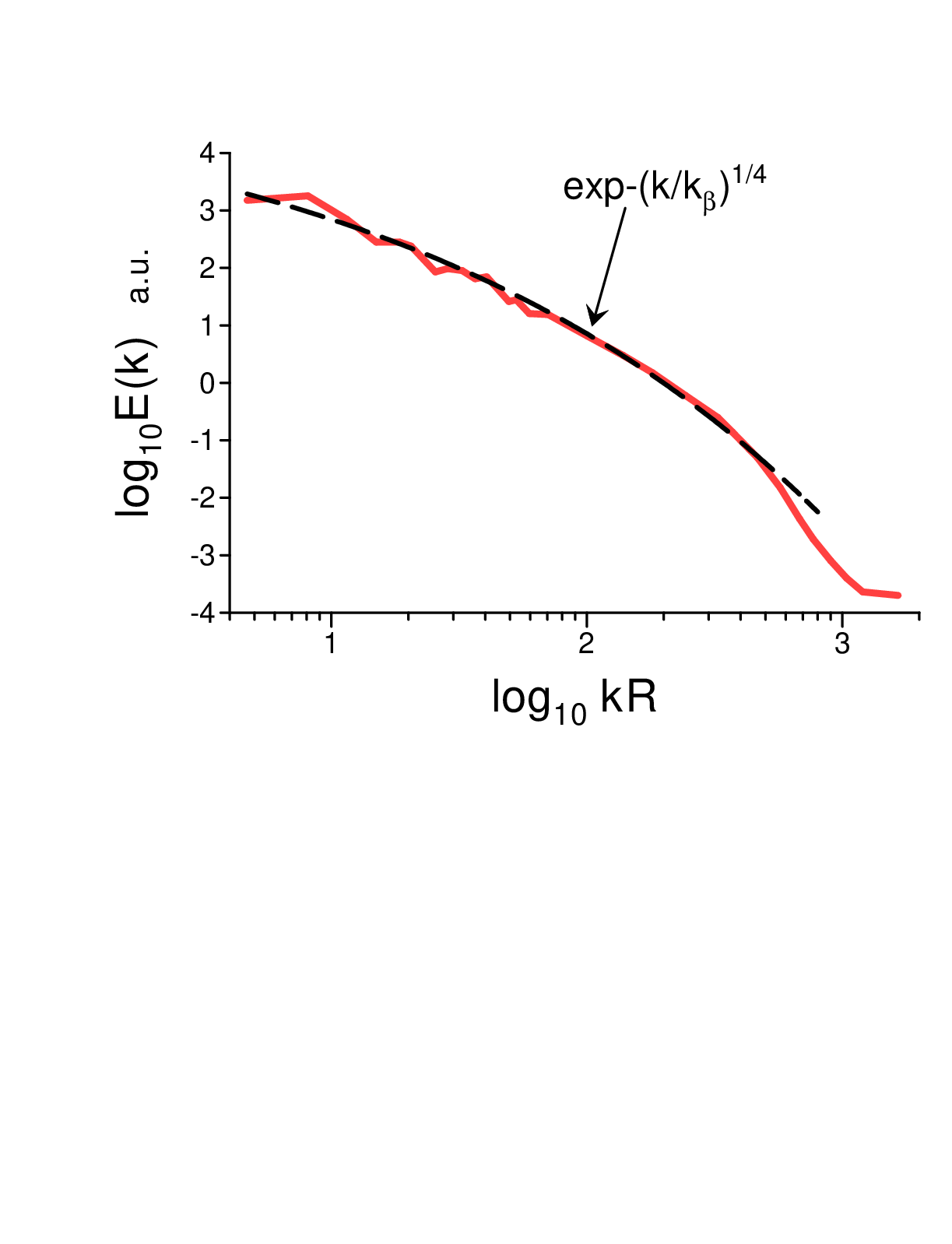} \vspace{-4.5cm}
\caption{Fourier power spectrum of the B-mode map for the Galactic Orion-Eridanus region.} 
\end{figure}
\begin{figure} \vspace{-0.5cm}\centering 
\epsfig{width=.65\textwidth,file=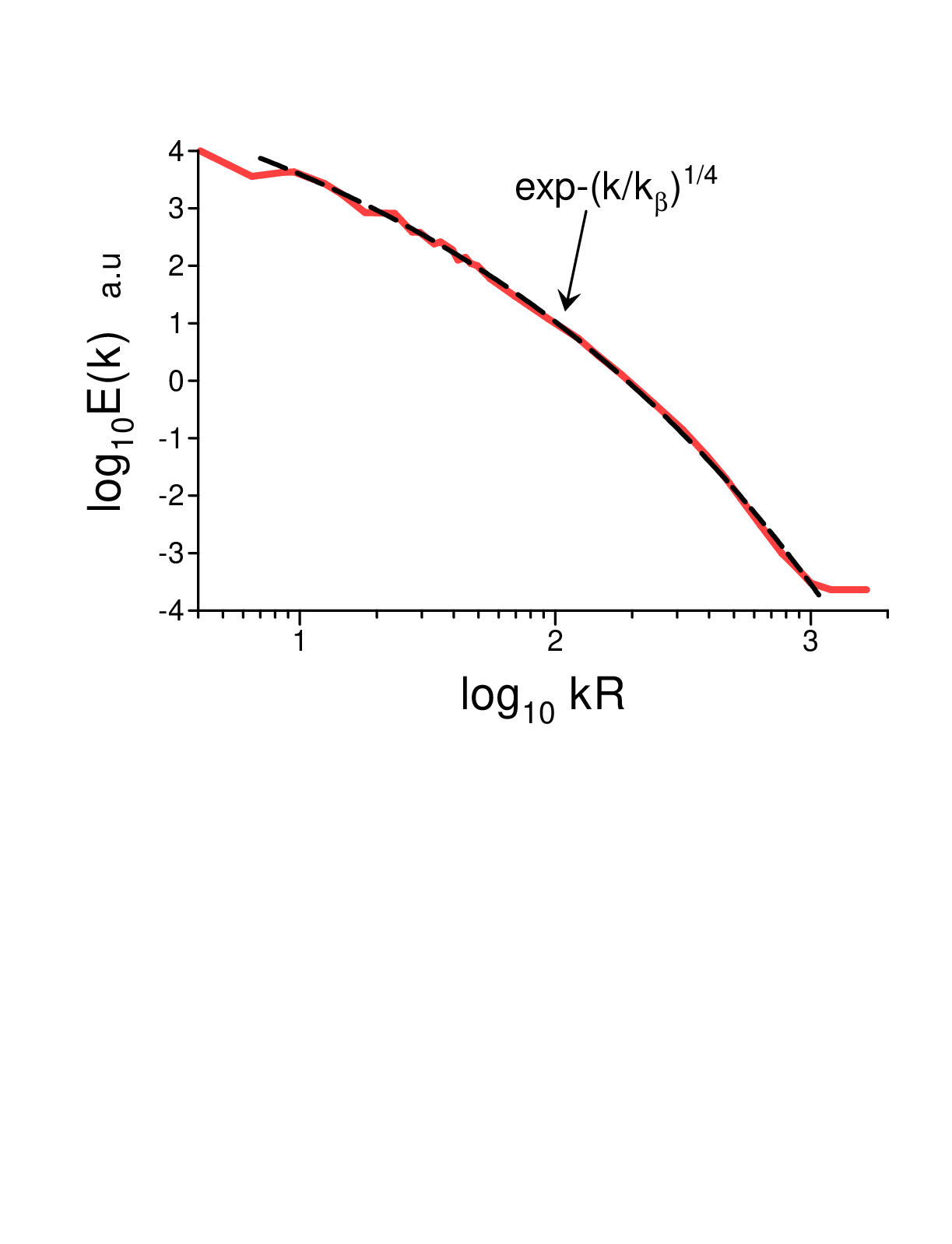} \vspace{-4.5cm}
\caption{Fourier power spectrum of the B-mode map for the Galactic G353-34 region.} 
\end{figure}
   It should be noted that in this case the synchrotron emission can be affected by the Faraday rotation (induced by the chaotic/turbulent magnetoionic plasma) along the line of sight (\citealt{rob}).\\
  
  Figures 10 and 11 show the Fourier power spectra of the B-mode maps for the Galactic Orion-Eridanus and G353-34 regions respectively. The spectral data were taken from Figs. 
12b and 13b of the Ref. (\citealt{rob}).  \\

  The dashed curves in Figs. 10 and 11 indicate the best fit by the stretched exponential spectral law Eq. (21). One can see that in this case as well the magnetic field imposes its degree of randomization.\\
\begin{figure} \vspace{-2.2cm}\centering 
\epsfig{width=.65\textwidth,file=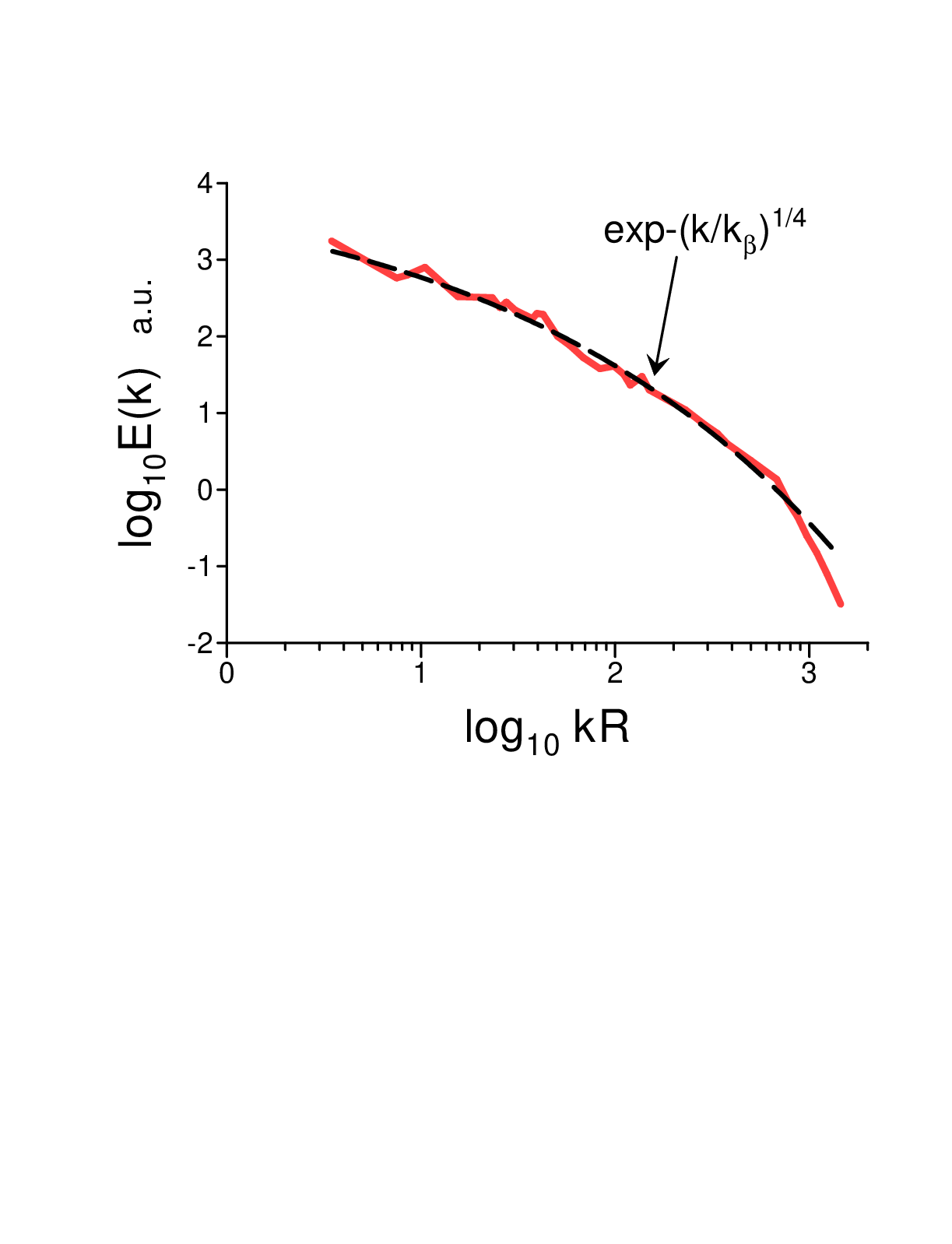} \vspace{-4.2cm}
\caption{Power spectrum of the B-mode component for the Parkes 2.4 GHz survey of synchrotron polarised emission of the Southern Galactic plane.} 
\end{figure}
\begin{figure} \vspace{-0.5cm}\centering 
\epsfig{width=.65\textwidth,file=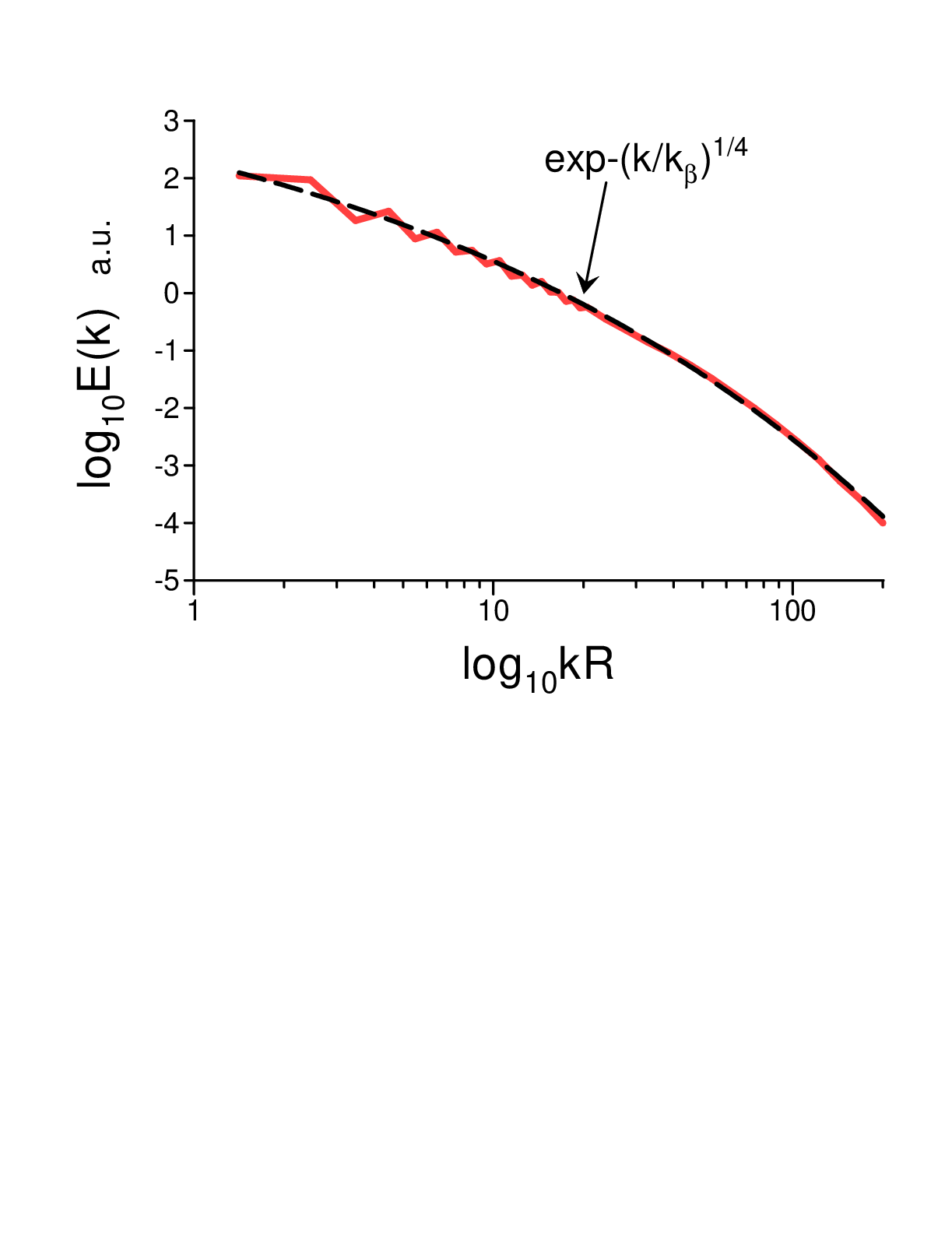} \vspace{-4.8cm}
\caption{Power spectrum for the `Haslam' 408 MHz all-sky survey of the synchrotron emission.} 
\end{figure}

  In a paper (\citealt{gia}) an analysis of the Parkes 2.4 GHz survey of synchrotron polarised emission of the entire Southern Galactic plane was performed. Figure 12 shows the power spectrum of the B-mode component. The spectral data were taken from Fig. 3 of the Ref. (\citealt{gia}). The dashed curve in Fig. 12 indicates the best fit by the stretched exponential spectral law Eq. (21). \\
  
  In a paper (\citealt{mer}) an analysis of the `Haslam' 408 MHz all-sky survey of the synchrotron emission was performed. Figure 13 shows the power spectrum obtained for this survey (the spectral data were taken from Fig. 3 of the Ref. \citealt{mer}). The point sources were subtracted and an averaging over different $m$-modes
(i.e. different lines-of-sight) was performed to obtain the original angular power spectrum. 

   The dashed curve in Fig. 13 indicates the best fit by the stretched exponential spectral law Eq. (21). One can see that in this case as well the magnetic field imposes its degree of randomization.\\

\section{Magnetic field of galaxy clusters and Faraday rotation maps} 

\begin{figure} \vspace{-2cm}\centering 
\epsfig{width=.57\textwidth,file=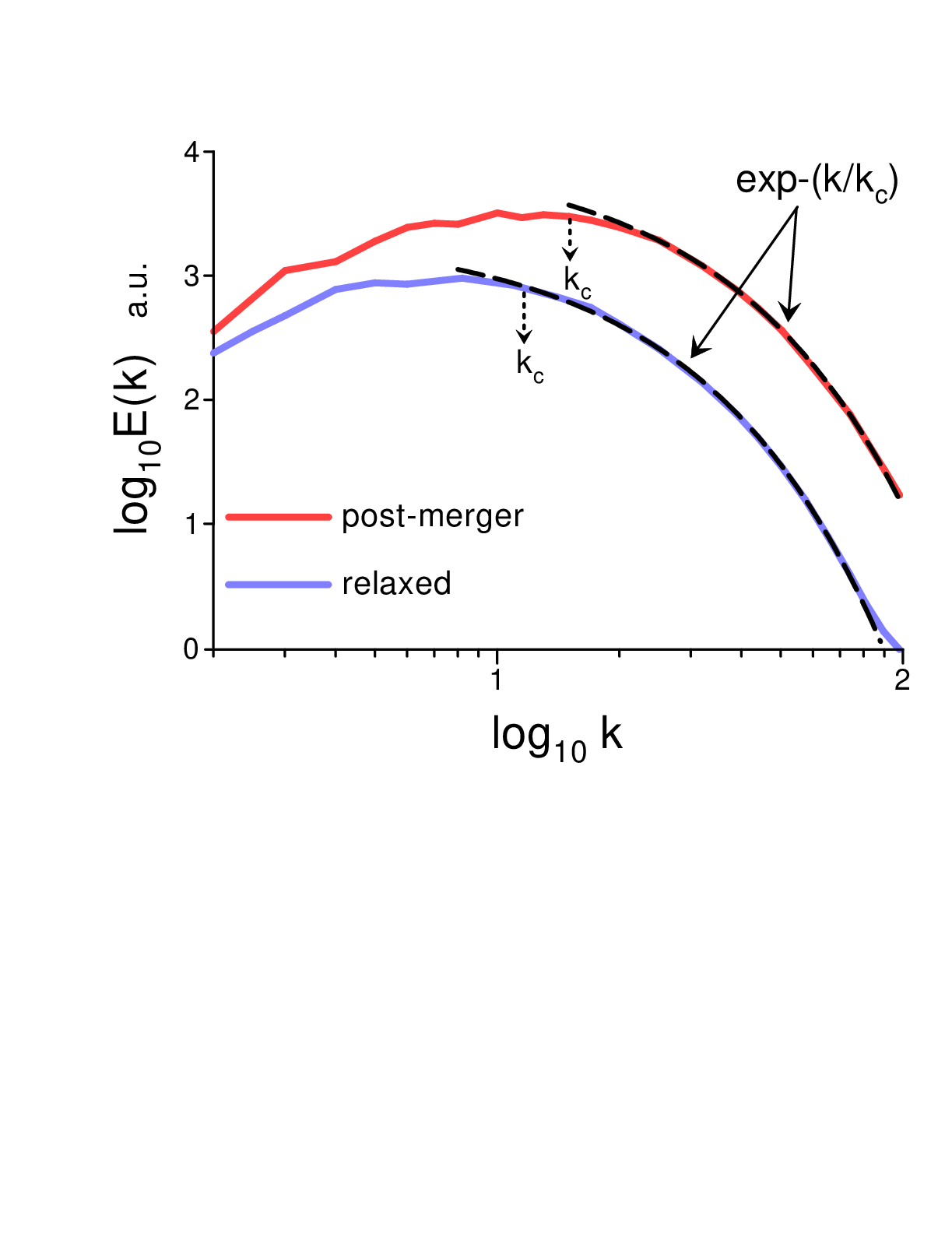} \vspace{-3.6cm}
\caption{Magnetic energy spectrum for the value of redshift $z=0$ (numerical simulations). } 
\end{figure}

 \subsection{Numerical simulations}
 
 {\bf A}.  In a recent paper  (\citealt{dom}) non-radiative cosmological magnetohydrodynamic simulations of galaxy clusters were performed to investigate the amplification of the primordial magnetic field via a small-scale dynamo driven by mergers. Dedner's formulation of the magnetohydrodynamic equations and $\Lambda$CMD cosmology were employed. \\
 
  Figure 14 shows the magnetic energy spectrum for $z=0$ ($z$ is the value of a redshift). The spectral data were taken from Fig. 4 of the paper (\citealt{dom}). The upper curve corresponds to a post-merger state with the maximal magnetic energy, whereas the bottom curve corresponds to a relaxed state. 
  
  The dashed curves in Fig. 14 indicate the best fit by the exponential spectral law Eq. (2) (i.e. the deterministic chaos).\\
  
 {\bf B}.  In a paper (\citealt{nak}) spectra of the magnetic energy and corresponding spectra of the Faraday rotational measure maps were computed in the framework of the standard (collisional) magnetohydrodynamics and in the framework of a collisionless MHD model (which is expected to be more appropriate to the magnetized and weakly collisional intracluster plasma). This model is based on the dynamic equations
\begin{equation}
 \frac{\partial\rho}{\partial t} + \nabla \cdot \left( \rho {\bf u} \right)  =  0, 
 \end{equation}
 \begin{equation}
 \frac{\partial\left(\rho\bf{u}\right)}{\partial t} + \nabla \cdot \left[ \rho\bf{u}\bf{u} + \left( \rm{\bf{P}} + \frac{B^2}{8\pi} \right) \rm\bf{I} - \frac{\bf{B}\bf{B}}{4\pi} \right]  = \bf{f}, 
  \end{equation}
 \begin{equation}
 \frac{\partial\bf{B}}{\partial t} - \nabla \times \left( \bf{u} \times \bf{B} \right)  =  0,
 \end{equation} 
where the pressure tensor is
\begin{equation}
 \rm{\bf{P}} = p_\perp {\bf I} + (p_\parallel - p_\perp)\hat{b}\hat{b}, \quad
 \end{equation}
here $p_{\perp} = c^2_{\perp} \rho $, $p_{\parallel} = c^2_{\parallel} \rho$, $\hat{b} = \bf{B}/|\bf{B}|$, and  $c_{\perp}$ and $c_{\parallel}$ are the sound
speeds perpendicular and parallel to the magnetic field $\bf{B}$, correspondingly.\\

  The authors of the paper (\citealt{nak}) reported some indications of a small-scale (fluctuation) dynamo in work for their models. \\

\begin{figure} \vspace{-2.2cm}\centering 
\epsfig{width=.6\textwidth,file=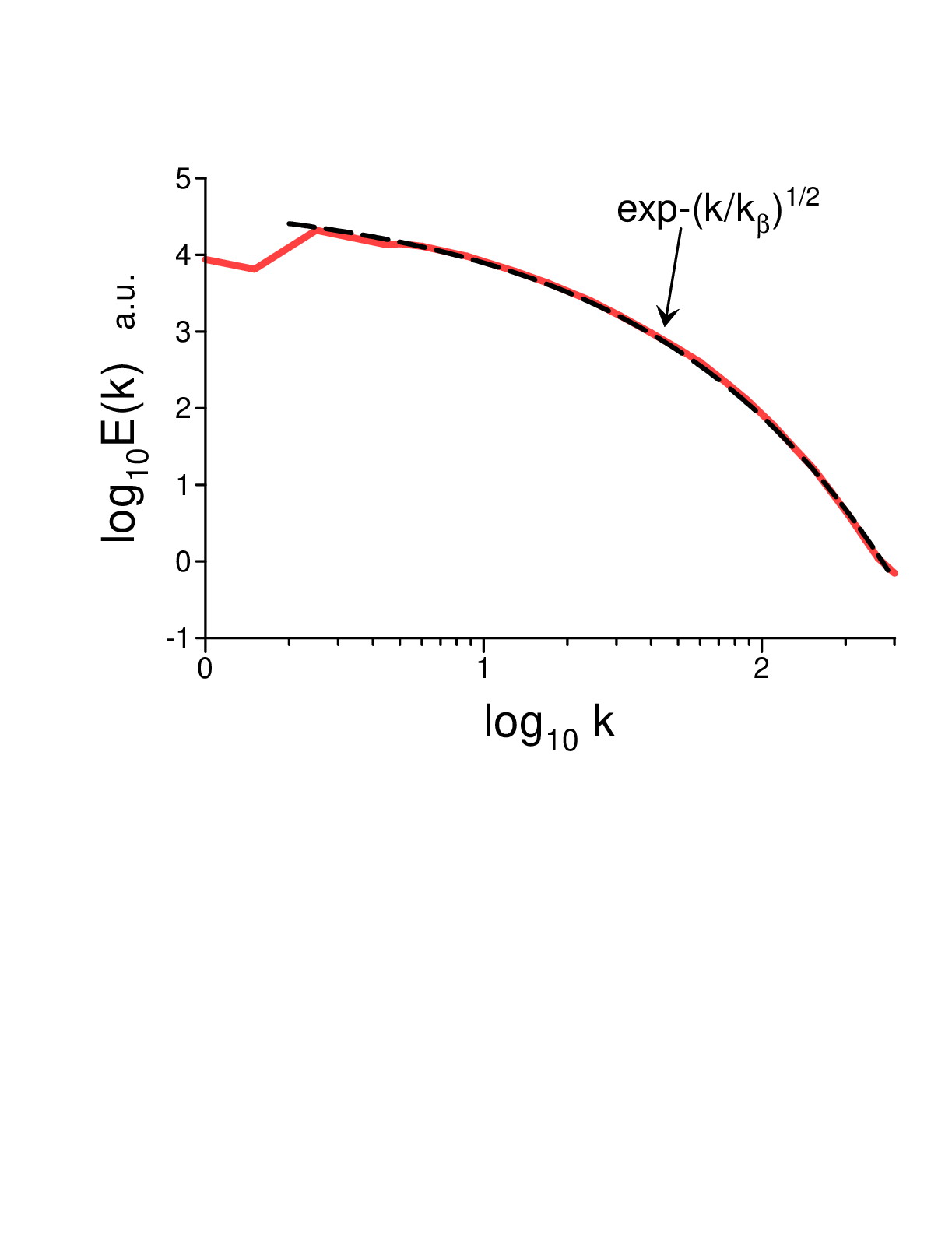} \vspace{-3.9cm}
\caption{Magnetic energy spectrum computed for the standard magnetohydrodynamics (numerical simulations). } 
\end{figure}
\begin{figure} \vspace{-0.5cm}\centering 
\epsfig{width=.6\textwidth,file=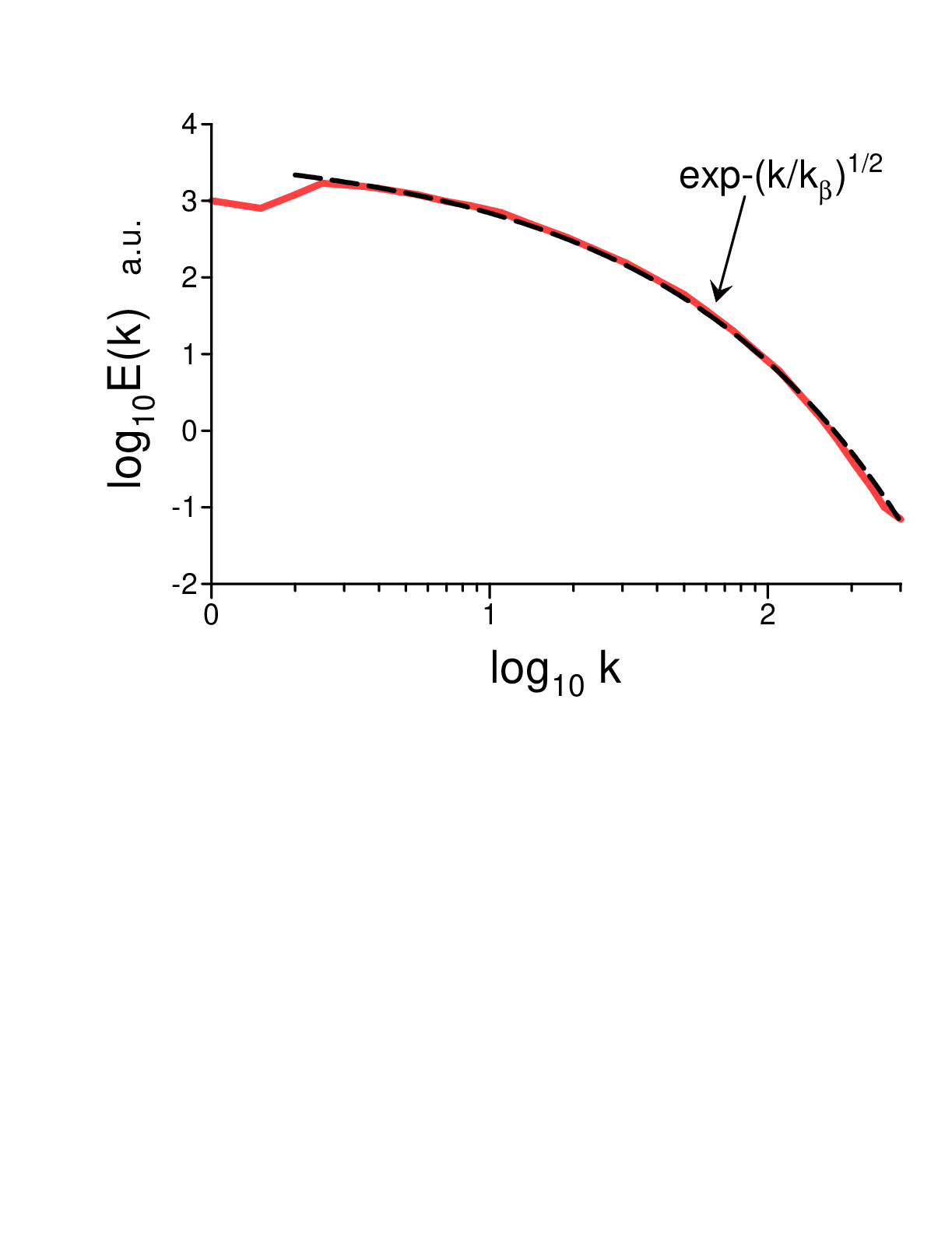} \vspace{-4.4cm}
\caption{Magnetic energy spectrum computed for the collisionless magnetohydrodynamics (numerical simulations).} 
\end{figure}
\begin{figure} \vspace{-2cm}\centering 
\epsfig{width=.6\textwidth,file=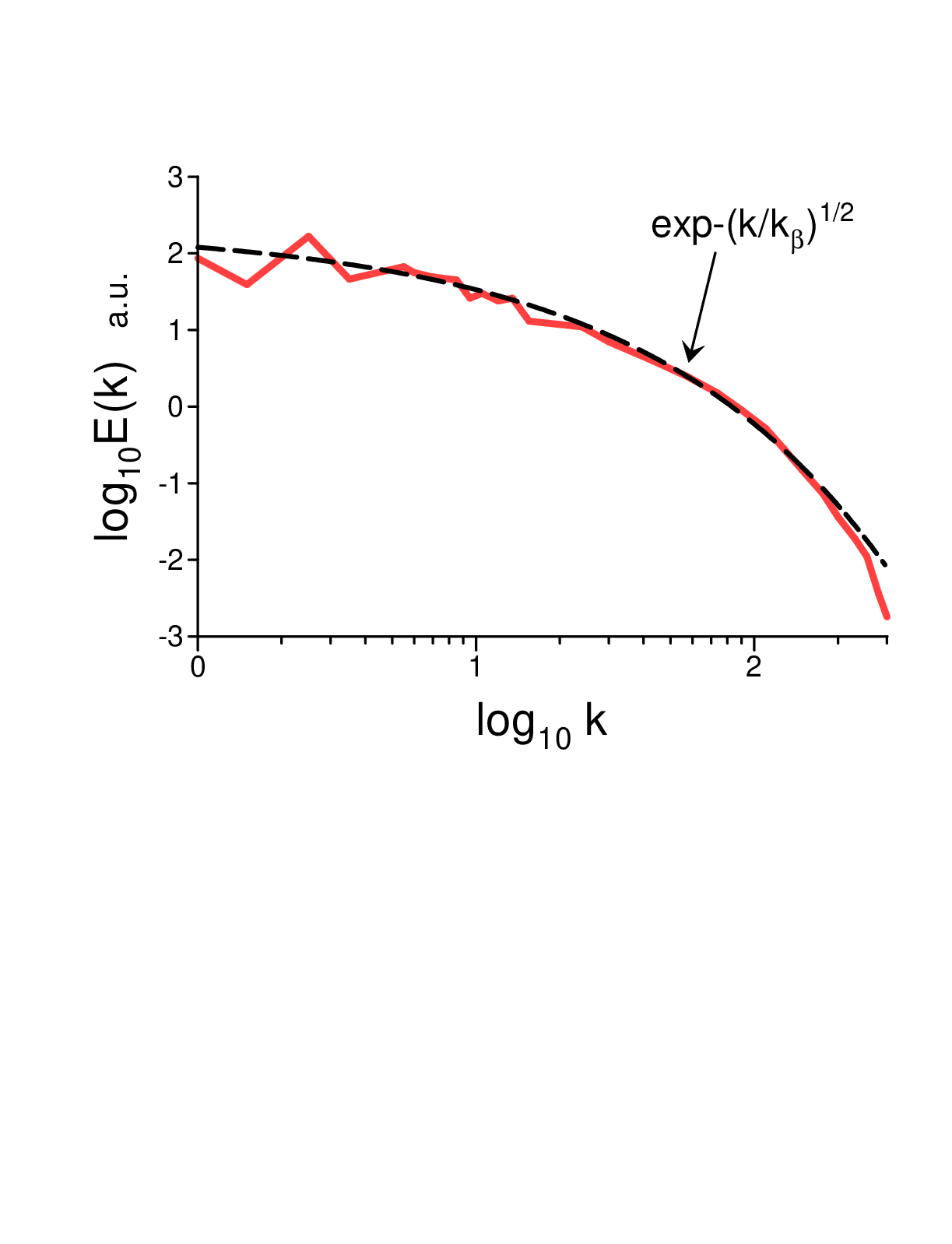} \vspace{-4cm}
\caption{Power spectra of the Faraday rotational measure map computed for the standard magnetohydrodynamics (numerical simulations). } 
\end{figure}
\begin{figure} \vspace{-0.5cm}\centering 
\epsfig{width=.6\textwidth,file=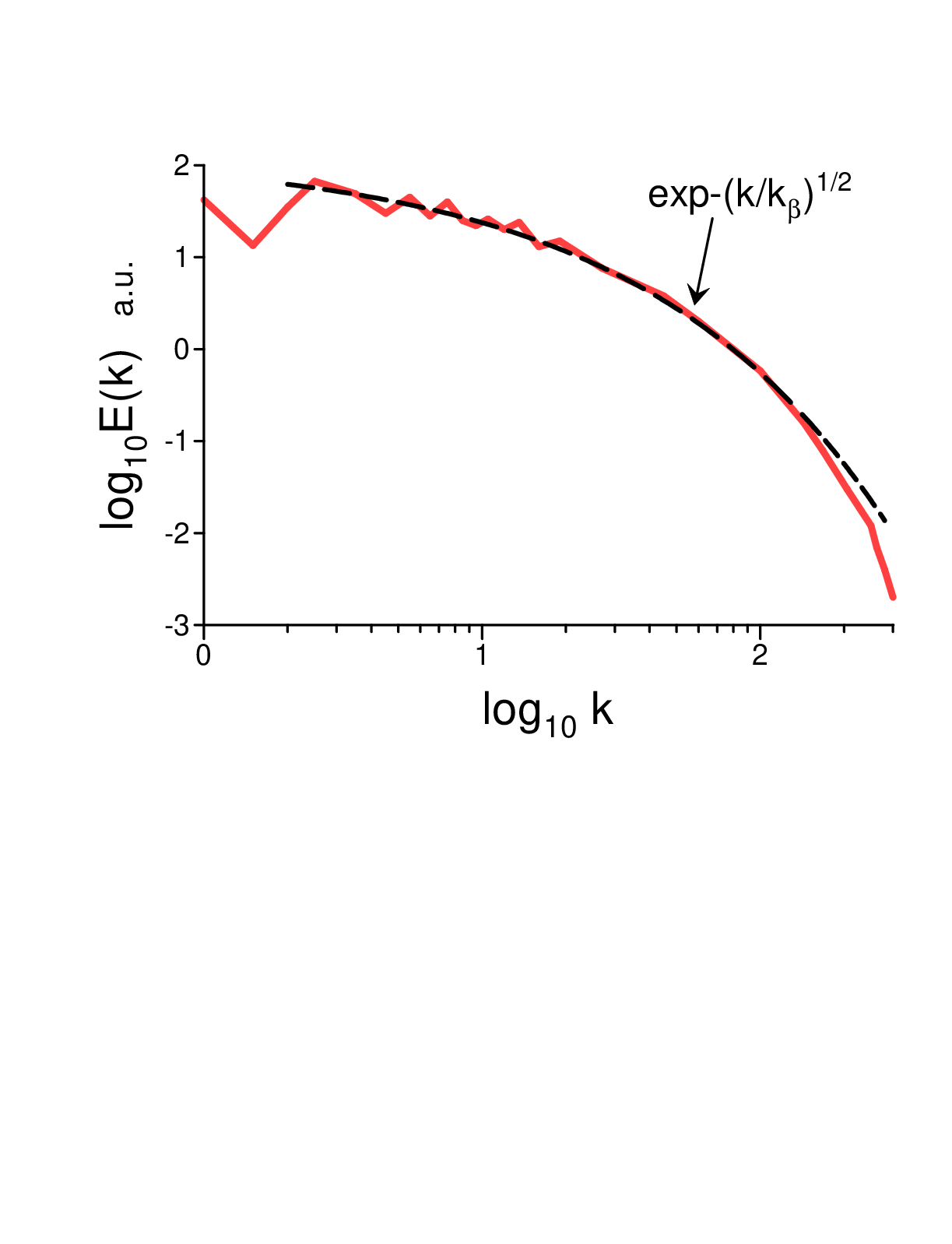} \vspace{-4.15cm}
\caption{Power spectra of the Faraday rotational measure map computed for the collisionless magnetohydrodynamics (numerical simulations). } 
\end{figure}

    Figure 15 shows the spectrum of magnetic energy computed for the standard magnetohydrodynamics with a weak uniform external magnetic field $B_{ext} = 0.1$ and sound speed $c_{snd} =0.1$ (in terms of \citealt{nak}). The spectral data were taken from Fig. 6a of the paper (\citealt{nak}). The dashed curve indicates the best fit by the stretched exponential Eq. (12).\\
    
    Figure 16 shows the spectrum of magnetic energy computed for the collisionless MHD model with a weak uniform external magnetic field $B_{ext} = 0.1$ and $c_{\parallel} = 0.1$, $c_{\perp} = 0.05$. The spectral data were taken from Fig. 6b of the paper (\citealt{nak}). The dashed curve indicates the best fit by the stretched exponential Eq. (12).\\
    
    One can see that for both models interpretation of the spectral data for magnetic energy using the notion of the distributed chaos dominated by magnetic helicity Eq. (12) is in good agreement with the results of the numerical simulations.\\
    
    Figures 17 and 18 show the power spectra of the Faraday rotational measure maps corresponding to Figs 15 and 16 respectively. The spectral data were taken from Figs 12a,c of the paper (\citealt{nak}). The dashed curves indicate the best fit by the stretched exponential Eq. (12).\\ 
    
    One can see that also in this case the magnetic field imposes its level of randomization on the Faraday maps (cf Subsection 4.2). Therefore the observational Faraday rotation measure maps could be used to obtain this information about the magnetic fields (see also below).

 \subsection{Observations}
 
\begin{figure} \vspace{-2cm}\centering 
\epsfig{width=.6\textwidth,file=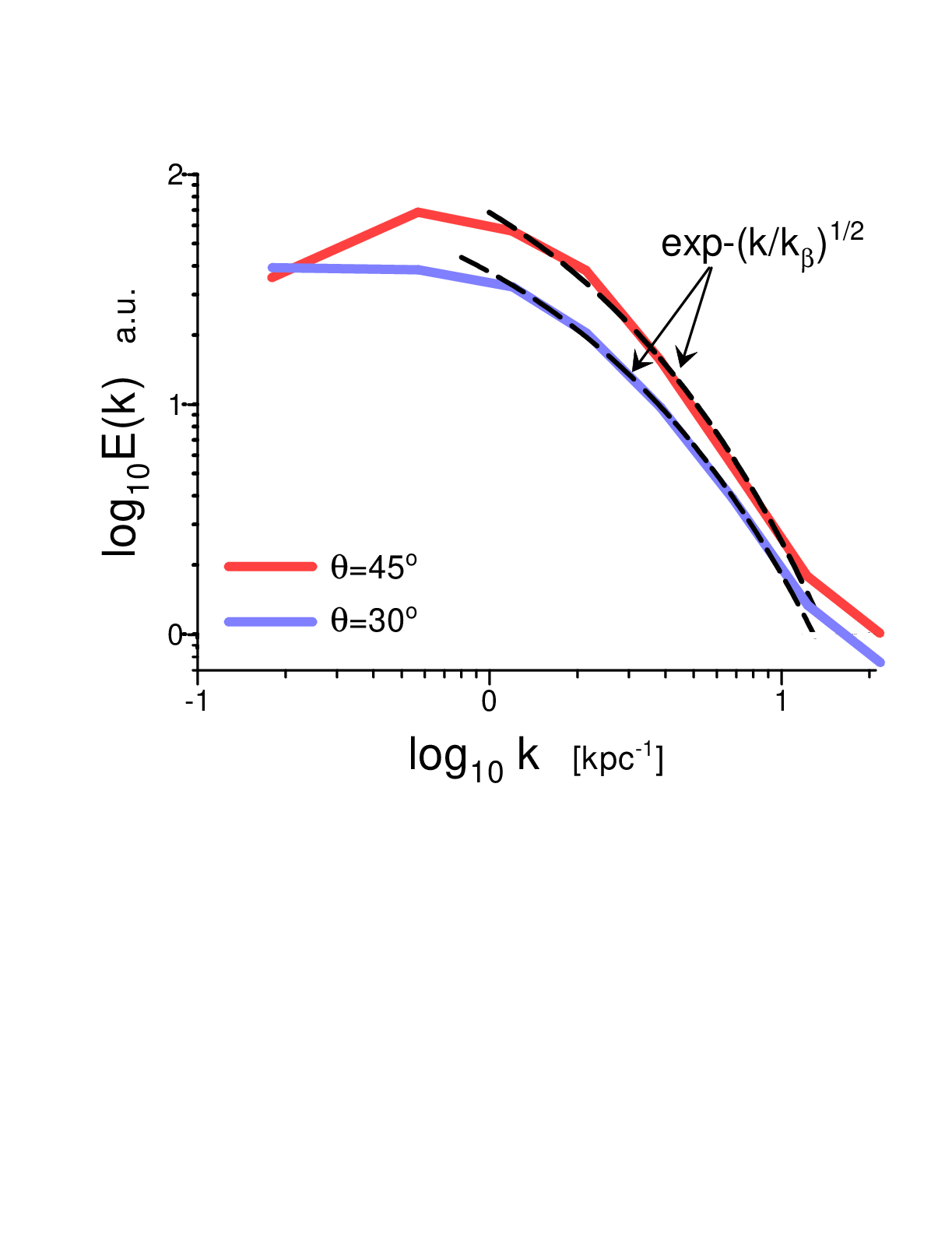} \vspace{-3.7cm}
\caption{Intracluster magnetic power spectra for the northern radio lobe of Hydra A for two values of the inclination angles (between the line of sight and the northern lobe) $\theta = 45^o$ (top) and  $\theta = 30^o$ (bottom).  } 
\end{figure}
\begin{figure} \vspace{-0.5cm}\centering 
\epsfig{width=.54\textwidth,file=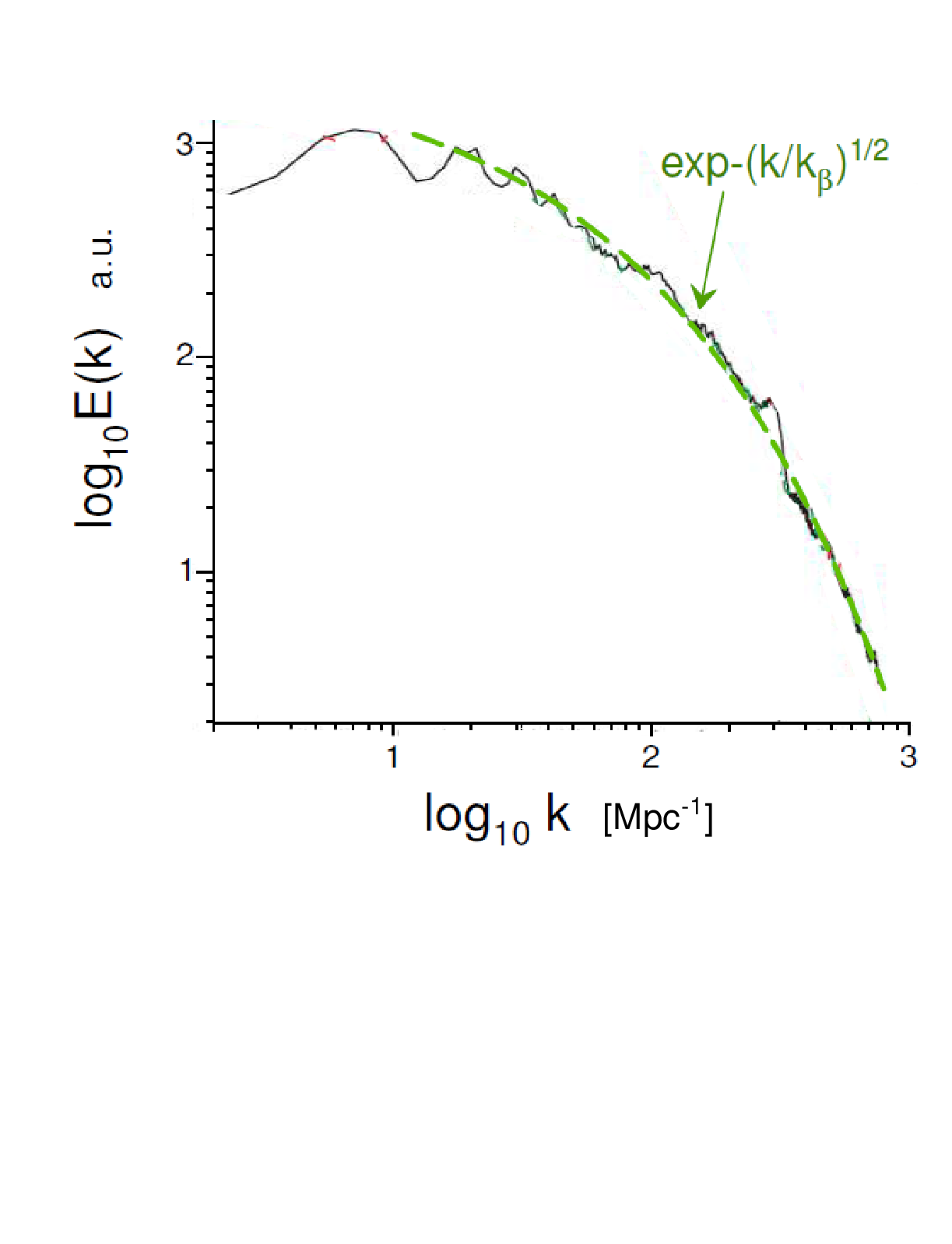} \vspace{-2.8cm}
\caption{Radially averaged power spectrum of a Faraday rotation measure map for the galaxy cluster A119 .} 
\end{figure}

    In a paper (\citealt{ve}) the Faraday rotation maps of the northern radio lobe of Hydra A (a cool core galaxy cluster) were analyzed with a Bayesian maximum likelihood analysis to infer the intracluster magnetic field power spectrum. Figure 19 shows the magnetic power spectra for two values of the inclination angles (between the line of sight and the northern lobe) $\theta = 45^o$ (top) and  $\theta = 30^o$ (bottom).  The spectral data were taken from Fig. 7 of the paper (\citealt{ve}).
    
    The dashed curve indicates the best fit by the stretched exponential Eq. (12) (cf previous section).\\ 
    
    Figure 20 shows the radially averaged power spectrum of a Faraday rotation measure map for the galaxy cluster A119 (the spectral data were taken from Fig. 10d of paper \citealt{mur}). This galaxy cluster is located in a region with a rather low Galactic rotation measure and consists of three extended radio galaxies. These galaxies are located at different projected distances (170, 453, and 1515 kpc) from the cluster center and the radio sources are highly polarized.\\
    
    The dashed curve indicates the best fit by the stretched exponential Eq. (12) (cf previous Section).\\
   
\section{Conclusions}
  
  The above-considered results obtained in numerical simulations (models) and inferred from the observations of the Faraday rotation and synchrotron emission maps show that magnetic helicity dominates the chaotic/turbulent dynamics (directly or through the Kolmogorov phenomenology) in the internal accretion disk plasma around the central black hole and in the global Galactic plasma's magnetic field, electron density, and the above-mentioned maps. The magnetic field imposes its level of randomization on the plasma's electron density and the Faraday rotation and synchrotron emission maps. Analogous conclusions can also be true for galaxy intracluster plasmas.\\
  
   The dominated by magnetic helicity randomization occurs even in the case of zero (or negligible) net helicity due to the spontaneous breaking of the local reflectional symmetry (intrinsic to the chaotic/turbulent plasma's dynamics).\\
   
  The applicability of the above-suggested approach to the kinetic scales (relevant to the internal Galactic disk's collisionless plasma) has also been established. \\
  
   Despite considerable differences in the physical parameters and scales, the results of the numerical simulations (models) related to the randomization phenomenon are in quantitative agreement with the observations of the plasmas of galaxies and galaxy clusters.

\section{Acknowledgments }

  I thank H.K. Moffatt, A. Pikovsky, and J.V. Shebalin for stimulating discussions.

\end{document}